\documentclass{aa}
\usepackage[varg]{txfonts}
\usepackage{hyperref} 
\usepackage{graphicx,caption}

\graphicspath{{./}{figures/}}
\usepackage{color}

\DeclareUnicodeCharacter{2000}{}

\newcommand\hmpc{\mathrm{h^{-1} Mpc}}

\begin{document}

\title{Sociology and hierarchy of voids:
A study of seven nearby CAVITY galaxy voids and their dynamical CosmicFlows-3 environment}

\author{H.M.~Courtois\thanks{helene.courtois@univ-lyon1.fr}\inst{\ref{inst1}}\and 
R.~van de Weygaert\inst{\ref{inst2}}\and
M.~Aubert\inst{\ref{inst1}}\and
D.~Pomarède\inst{\ref{inst3}}\and
D.~Guinet\inst{\ref{inst1}}\and
J.~Dom\'{i}nguez-G\'{o}mez\inst{\ref{inst4}}\and 
E.~Florido\inst{\ref{inst4},\ref{inst5}}\and
L.~Galbany\inst{\ref{inst6},\ref{inst7}}\and 
R.~Garc\'{i}a-Benito\inst{\ref{inst8}}\and
J.M.~van der Hulst\inst{\ref{inst2}}\and
K.~Kreckel\inst{\ref{inst9}}\and 
R.E.~Miura\inst{\ref{inst4},\ref{inst10}}\and
I. P\'erez\inst{\ref{inst4},\ref{inst5}}\and 
S.~Planelles\inst{\ref{inst11},\ref{inst12}}\and 
V.~Quilis\inst{\ref{inst11}, \ref{inst12}}\and 
J.~Rom\'{a}n\inst{\ref{inst2},\ref{inst13},\ref{inst14}}\and
M.~S\'{a}nchez-Portal\inst{\ref{inst15}}
}

\institute{Universit\'e Claude Bernard Lyon 1, IUF, IP2I Lyon, 69622 Villeurbanne, France\label{inst1}
\and Kapteyn Astronomical Institute, University of Groningen, PO Box 800, 9700 AV Groningen, The Netherlands\label{inst2}
\and IRFU, CEA Universit\'e Paris-Saclay, 91191 Gif-sur-Yvette, France\label{inst3}
\and Departamento de F{\'i}sica Te{\'o}rica y del Cosmos, Campus de Fuente Nueva, Edificio Mecenas, Universidad de Granada, E-18071, Granada, Spain\label{inst4}
\and Instituto Carlos I de F\'{i}sica Te\'{o}rica y Computacional, Facultad de Ciencias, E-18071 Granada, Spain\label{inst5}
\and Institute of Space Sciences (ICE, CSIC), Campus UAB, Carrer de Can Magrans, s/n, E-08193 Barcelona, Spain\label{inst6}
\and Institut d’Estudis Espacials de Catalunya (IEEC), E-08034 Barcelona, Spain\label{inst7}
\and Instituto de Astrof\'{i}sica de Andaluc\'{i}a - CSIC Glorieta de la Astronom\'{i}a E-18008, Granada, Spain\label{inst8}
\and Astronomisches Rechen-Institut, Zentrum f\"{u}r Astronomie der Universit\"{a}t Heidelberg, M\"{o}nchhofstrasse 12-14, D-69120 Heidelberg, Germany\label{inst9}
\and National Astronomical Observatory of Japan, National Institutes of Natural Sciences, 2-21-1 Osawa, Mitaka, Tokyo 181-8588,
Japan\label{inst10}
\and Departamento de Astronom\'{i}a y Astrof\'{i}sica, Universidad de Valencia, c/ Dr. Moliner n  50, 46100 - Burjassot (Valencia), Spain\label{inst11}
\and Observatori Astronòmic, Universitat de València, E-46980 Paterna (València), Spain\label{inst12}
\and Departamento de Astrof\'{\i}sica, Universidad de La Laguna, E-38206, La Laguna, Tenerife, Spain\label{inst13}
\and Instituto de Astrof\'{\i}sica de Canarias, c/ V\'{\i}a L\'actea s/n, E-38205, La Laguna, Tenerife, Spain\label{inst14}
\and Institut de Radioastronomie Millim\'etrique (IRAM), Av. Divina Pastora 7, Local 20, 18012 Granada, Spain\label{inst15}
}

\date{Received A\&A Nov 29, 2022 - AA/2022/45578 / Accepted March 3rd, 2023 }

\abstract
{The present study addresses a key question related to our understanding of the relation between void galaxies and their environment: the relationship between luminous and dark matter in and around voids.}
{To explore the extent to which local Universe voids are empty
of matter, we study the full (dark+luminous) matter content of seven nearby cosmic voids that are fully contained within the CosmicFlows-3 volume. }
{We obtained the matter-density profiles of seven cosmic voids using two independent methods. These were built from the galaxy redshift space two-point correlation function in conjunction with peculiar velocity gradients from the CosmicFlows-3 dataset.}
{ The results are striking, because when the redshift survey is used, all voids show a radial positive gradient of galaxies, while based on the dynamical analysis, only three of these voids display a clear underdensity of matter in their center.}
{ This work constitutes the most detailed observational analysis of voids conducted so far, and shows that void emptiness should be derived from dynamical information. From this limited study, the Hercules void appears to be the best candidate for a local Universe pure "pristine volume", expanding in three directions with no dark matter located in that void.}


\titlerunning{Local universe voids emptiness}
\authorrunning{Courtois et al.} 
\maketitle
\section{Introduction}

\noindent Galaxies are not evenly distributed in space. Instead, they probe the underlying inhomogeneous dark matter distribution.
On megaparsec scales, matter and galaxies are organized in a weblike network called the cosmic web \citep{bond1996,cautun2014}.
Prominent elongated filaments define a pervasive structure that assembles most of the matter and galaxies in the Universe and form intergalactic transport channels along which mass is migrating towards the dense, compact clusters and the nodes of
the web. An equally outstanding aspect of the weblike cosmic mass distribution is the nearly~ empty void regions \citep[e.g.,][]{kirshner1981,lapparent1986,huchra2012}. These are enormous regions with sizes in the range of $20-50h^{-1}$ Mpc that are significantly less populated with galaxies than filaments and clusters. Voids are generally roundish in shape and occupy the major share of space \citep[see e.g.,][for reviews]{weygaert2011,weygaert2016},
assuming around $75\%-80\%$ of it \citep[e.g.,][]{cautun2014}.

The dominant voids in the cosmic matter distribution are manifestations of the cosmic-structure-formation process transiting to the nonlinear stage of evolution \citep{blumenthal1992,shethwey2004}. Their effective repulsive influence over the surroundings has  even been recognized in the CosmicFlows, which are peculiar velocity surveys of the Local Universe \citep{courtois2012,tully2014}.
The expansion of the voids makes them organizing elements of the large-scale matter distribution, meaning they play an essential role
in arranging matter concentrations into an all-pervasive cosmic network \citep[e.g.,][]{icke1984,shethwey2004,aragon2013}.

Many recent studies followed up on the realization that voids not only represent a key constituent of the cosmic-mass distribution,  but are also one of the most direct probes of global cosmology  \citep{goldberg2004,parklee2007,lavaux2011,bos2012,pisani2015,hamaus2016,cai2015,perico2019}. Of particular interest is the realization that their structure, morphology, and dynamics reflect the nature of dark energy,
dark matter, and of the possibly non-Gaussianity of the primordial perturbation field \citep[see][for a review]{pisani2020}. The effects of
dark energy and possible modifications of General Relativity manifest themselves more prominently in the low-density interior of voids.

The interior of voids also offer a unique testing ground for studying environmental influences on galaxy formation and evolution
\citep{peebles2001,rojas2004,rojas2005,kreckel2011,kreckel2012}. The low-density interior of a void is a largely pristine cosmic environment
that still retains the memory of the initial conditions of the Universe soon after the big bang, and is unaffected by virialization or other effects related to gravitational collapse.
Equivalent to a lower $\Omega_m$
universe \citep{goldberg2004}, galaxies in voids are expected to evolve more slowly and have a "calmer" merging history \citep[also see][]{lackner2012}. As a result, they appear to have significantly different properties from average field galaxies.

In general, void galaxies are small, faint, and blue galaxies \citep{kreckel2011} that appear to reside in a more youthful state of
star formation than galaxies in denser environments \citep{beygu2016,dominguezgomez2022}. Nonetheless, while these trends appear to
be quite general, controversy persists in the literature as to whether or not galaxies in voids genuinely differ in their internal properties
from similar objects in denser regions. For example, while \cite{rojas2005} found evidence for a signficantly higher specific star formation rate for void galaxies, other studies \citep[e.g.,]{beygu2016} did not find evidence that void galaxies have star formation rates in excess of what
might be expected for their small mass. It is a well-established fact that there are systematic differences in the masses of halos and
galaxies residing in different cosmic-web environments: with respect to the generic filament environment of galaxies, the void galaxy population
mass function is shifted to lower masses and lower number densities \citep[see e.g.,]{cautun2014,punya2019,hellwing2021}. A key
question pertains to whether the observed systematic differences between galaxies in voids, filaments, walls, and clusters are only due to the differences
in their mass, with other noticeable differences solely a direct consequence of this, or there are environment-specific factors
at play \citep[see e.g.,][]{borz2017,hellwing2021}. In a recent study, \cite{goh2019} argued in favor of mass being the sole determining factor. On the
other hand, specifically in the case of void galaxies, \citet{peebles2001}, in his seminal study on the "void phenomenon", pointed out the
unexpected low abundance of low-mass and dwarf galaxies in voids.  This appears to be difficult to reconcile with mere mass scaling within
standard LCDM cosmology and may be the strongest indication of specific void environmental processes. Indeed, there are many additional
environmental factors and processes that one might expect to contribute to the outcome of the galaxy-formation process. For example, a prominent environmental influence is that of the different external tidal forces exerted on a forming halo and galaxy \citep[see e.g.,][]{weybab1994, hahn2007, yan2013, borz2017, paranjape2018, verza2022}. 

With the purpose of investigating systematic differences between void galaxies and galaxies in denser environments, 
the Void Galaxy Survey (VGS) \citep{kreckel2011,kreckel2012,beygu2016,beygu2017} carried out a multi-wavelength study of
about 60 void galaxies. Each galaxy was selected from the deepest interior regions of identified voids 
in the The Sloan Digital Sky Survey (SDSS) redshift survey on the basis of a geometric--topological watershed technique \citep{platen2007}, with no
a priori selection of intrinsic properties of the void galaxies. These authors studied the gas content, star formation
history, and stellar content  in detail, as well as the kinematics and dynamics of void galaxies and their companions. One of the most tantalizing
findings of the VGS is the possible evidence for cold-gas accretion in some of the most interesting objects, amongst which are a 
polar ring galaxy and a filamentary configuration of void galaxies.

The Calar Alto Void Integral-field Treasury surveY (CAVITY\footnote{https://CAVITY.caha.es/}; \citep{Perez2023}) is the sequel to the VGS and extends the scope of the
observational study of the void galaxies. The CAVITY project concentrates  on the determination of the influence of the cosmic environment on
galaxy formation and the mass-assembly history of void galaxies, and in particular on the drivers of galaxy transformation in voids. Insight into the dynamical state of the voids in which galaxies are located
and, in particular, a sound understanding of the local relation between their
dark matter and luminous matter contents  are of utmost
importance for a reliable interpretation of measurements of void galaxies.\\
 
The present study addresses a key open question related to our understanding of the link between void galaxies and their environment, namely
the relationship between luminous and dark matter in and around voids. To this end, we seek to relate the galaxy distribution
in and around a sample of nearby voids with a dynamical study of these voids.  Our dynamical study is based on peculiar velocity measurements provided by the CosmicFlows-3 survey (CF3 ; \citep{2016AJ....152...50T}). We use this information to analyze the emptiness of these seven nearby (low-redshift) CAVITY voids, and
to assess the corresponding relationship to the void galaxy environment. 

\section{Data: CAVITY and CosmicFlows-3}

The CAVITY project targets void member galaxies spread across
15 singular voids. The seven voids analyzed here are taken from the CAVITY void sample on the basis of them being located in the CosmicFlows-3 reconstructed volume of the
Local Universe.   We use the number label that was given by \citep{2012MNRAS.421..926P}  in order to refer to these seven voids in the present article, namely 355, 439, 474, 487, 727, 738, and 941.

\subsection{CAVITY}
Calar Alto Observatory selected the CAVITY project as one of the three Legacy projects that will define the Calar Alto Observatory science and technology horizon for the coming years. The CAVITY project main goals are to determine the influence of the environment on the mass assembly of void galaxies,
to establish how galaxy formation depends on the larger-scale environment, and to identify the main driver of galaxy transformation in voids.

To carry out this detailed study, the CAVITY galaxy sample was selected using the Catalogue of Cosmic Voids based on SDSS DR7 data \citep{2012MNRAS.421..926P}. The selected redshift range is between 0.005 and 0.05 in order to obtain photometric and spectroscopic data reaching high-angular-resolution and faint galaxies in order to have a representative sample
of galaxies within voids. To obtain a precise characterization of the voids, they should be fully included in the survey and should contain at least 20 galaxies spread at various void-centric distances. Voids located near the edges of the SDSS footprint were eliminated because their centers and geometries could not be properly assessed. After a careful statistical characterisation of the remaining voids, a subsample of voids was selected that spans the largest ranges of effective
radius, number of galaxies, and volume number density of galaxies. This defines a mother sample of around 3\,000 galaxies. A subsample of the order of 200-300 galaxies is being observed with the PMAS instrument \citep{2005PASP..117..620R} on the 3.5m telescope of Calar Alto observatory.

\subsection{CosmicFlows-3}
In order to assess the dynamics of the voids in the CAVITY sample, we use the matter distribution implied by the peculiar galaxy velocities in the
CosmicFlows-3 (CF3) catalog. The third release of the CosmicFlows catalog compiles about 18\,000 measurements of galaxy distances and provides
a corresponding cosmography on a three-dimensional grid in supergalactic coordinates, which is used in this article. The recently released fourth version of the
CosmicFlows catalog (CF4) delivers about 56\,000 galaxies and about 1\,000 Type Ia supernovae distance measurements \citep{2023ApJ...944...94T}.

The combined measurements of galaxy luminosity distances and recessional velocities in the CF3 catalogue allows us to map the full matter overdensity field in
the local Universe out to $z<0.05$. The observational data are combined within an iterative forward modeling analysis \citep{2019MNRAS.488.5438G} within the LCDM paradigm of \cite{Planck2015}.
This analysis entails a comprehensive incorporation of local over- and underdensities and their associated peculiar (or gravitational) velocity fields.

Assuming that the mass fluctuations $\delta_m(\mathbf{x},t)$ reside in the linear regime, the full dark+luminous matter at position and time
($\mathbf{x},t$) is obtained on the basis of the reconstructed full 3D matter peculiar velocity field $\boldsymbol{v_m}$ by:
\begin{equation}
    \boldsymbol{\nabla.v_m}=-aHf(\Omega_{m})\delta_m(\mathbf{x},t)\,,
    \label{peculiar velocity growth index}
\end{equation}
in which $a$ is the scale factor of the Universe, $H$ is the Hubble expansion rate, $\Omega_m$ is the cosmological mass-density parameter, and $f(\Omega_{m})$ is the (linear) structure growth
rate \citep{peebles1980}. The growth rate $f(\Omega_{m})$ depends on the cosmological parameter $\Omega_{m}$ used for the computation.
The CF3 density field $\delta_m(\mathbf{x})$ is computed on a $256^3$ grid of size 500 $\mathrm{h^{-1}Mpc}$. This yields a resolution of about
$2 \mathrm{h^{-1}Mpc}$ per voxel (A voxel is the smallest resolved volume in a 3D grid. Its size is
the resolution of the grid.)

\section{Void analysis}
As we intend to compare the matter content of voids inferred from the galaxy redshift distribution to that computed from
the peculiar velocity dynamics, we need to follow void-identification procedures that allow us to trace voids in the two corresponding
situations. The first void-detection procedure seeks to trace voids in the discrete spatial distribution of galaxies.

\subsection{Void detection and identification in the SDSS galaxy redshift survey}

The catalogue of cosmic voids extracted from the SDSS DR7 galaxy sample was identified with the
\texttt{VoidFinder} procedure \citep{el-ad_voids_1997,2002ApJ...566..641H, 2004ApJ...607..751H, 2012MNRAS.421..926P}. In a first step, the algorithm classifies the galaxies of the sample into field and wall galaxies using the third nearest neighbour distance $d_3$ in relation to a threshold distance $d_{thr} = 6.3 \hmpc$. A field galaxy is a candidate for belonging to a void-like region ( $d_3$ > $d_{thr}$ ),  whereas a wall galaxy lies within a denser region, such as a filament or cluster  ( $d_3$ > $d_{thr}$ ). The wall galaxies are used to define the survey density grid. Starting from empty grid cells, empty spheres
are grown until they can be defined by four boundary wall galaxies. The largest spheres are taken to be voids. The smaller spheres with more than 10 $\%$ overlap are merged with the larger voids. Once these mergers are complete, the void centers are taken to be the centers of mass of the empty spheres defining a void. An in-depth description of the procedure is given in \citet{2002ApJ...566..641H} and the specific details of this DR7 sample are provided in \citet{2012MNRAS.421..926P}.

\subsection{Void characterization and void parameters}
The most prominent void properties are their size and density profiles.

\subsubsection{Void density profiles}
The major goal of our analysis is to compare the radial galaxy number density profile $\delta_g(r)$ and the
corresponding radial mass-density profile $\delta^{CF3}_{m,v} (r)$ of each of the voids in our sample and to decipher whether or not their exists a relationship between the two. We seek to establish how far galaxies in and
around the voids in our sample trace the underlying mass distribution inferred from the velocity field analysis in the CF3 catalog.

\bigskip
We computed the galaxy number density profile $\delta_g(r)$ of the  seven voids of our sample using redshift survey positions of the galaxies residing within them (with equation~\ref{xi}).
This profile is estimated by counting the number of SDSS galaxies in various shells of radii $r_i/R_v$ around the
center of each void. The profile computation involves 40 radial bins of width $\Delta r=0.125$~$R_v$, ranging over the interval $r_i/R_v$=0.0 to 5.0. 
The number counts of galaxies in the radial bins around the void centers are normalized by the number counts of a random galaxy sample that
follows the same angular coverage and radial selection function as the galaxy sample, as expressed formally in equation~\eqref{xi}, in which the
random sample consists of ten times more objects than the observed galaxy sample.

The resulting  radial density profile estimate $\delta_g(r)$ is equal to the radial correlation function $\xi(r)$ around a single void center: 
\begin{equation}
    \delta_{g} (r) := \xi(r) = \frac{n_R}{n_D}\frac{D_{vg}(r)}{R_{vg}(r)} - 1.
    \label{xi}
\end{equation}
In the above expression, we follow the Davis-Peebles estimator of the correlation function \citep{davispeebles1983}. The number of galaxies around a void
center at radius $r$ is given by $D_{vg}(r)$, while the number of galaxies around a void center in the random galaxy sample is given by $R_{vg}(r)$.
The ratio between the total number of data sample galaxies $n_D$ and that in the random sample $n_R$,  $n_R/n_D$, constitutes the required normalization.
Hence, in practice, the estimate of $\xi(r)$ is the galaxy number density contrast. 

\bigskip
The radial mass-density profiles $\delta^{CF3}_{m,v} (r/R_v)$ of each void are determined from the mass-density field $\delta_m$ reconstruction from
the CF3 survey. The mass-density field is represented on a $256^3$ grid of $2 \mathrm{h^{-1}Mpc}$ voxels. The radially averaged mass-density profile
$\delta^{CF3}_{m,v} (r_i/R_v)$ around voids measures the mean full matter (dark+luminous) density from the void center $r = 0$ to its outskirts
$r/R_v > 1$.

In order to  also include the environment of each individual void, the mass and galaxy number density profiles are computed from $r/R_v\,=\,0,\ldots,5$.  
The profile measures the average matter density in radial shells of width $\Delta r$, including the voxels $i$ with a radial distance in the range 
$r-\Delta r/2 < r_i < r+ \Delta r/2$. The profile is determined from the grid-based density field $\delta^{CF3}_{m,i}$ as follows:  
\begin{equation}
    \delta^{CF3}_{m,v} (r/R_v) = \frac{1}{N_{\mathrm{voxel}}} \sum_i \delta^{CF3}_{m,i}\,.
\label{delta_CF3}    
\end{equation}
Hence, the void matter density profile computed from the CF3 density grid is the average density in radial bins
of size $\Delta r = 5/40 = 0.125~R_v$, where $N_{\mathrm{voxel}}$ is the number of voxels found at a separation of $r/R_v$ and $\delta_i^{CF3}$ is the value of the CF3
density field in the voxel $i$ found at a separation of $r_i/R_v$.

The CF3 grid resolution is 2 $h^{-1}$Mpc, while the CAVITY voids ---which are identified using a galaxy number density field--- span an interval of mean effective radii of 15 $\mathrm{h^{-1}Mpc} < R_{v}< 25 \mathrm{h^{-1}Mpc}$.

\section{Voids and their large-scale environment}
\label{sec:environment}
In order to be able to interpret the results we obtain for the relation between the distribution of galaxies and the total mass in and around
voids, we carried out a visual inspection. This enables a direct study of the large-scale environment of the void sample. 

\begin{figure*}[htp]
    \centering
    \includegraphics[scale=0.4]{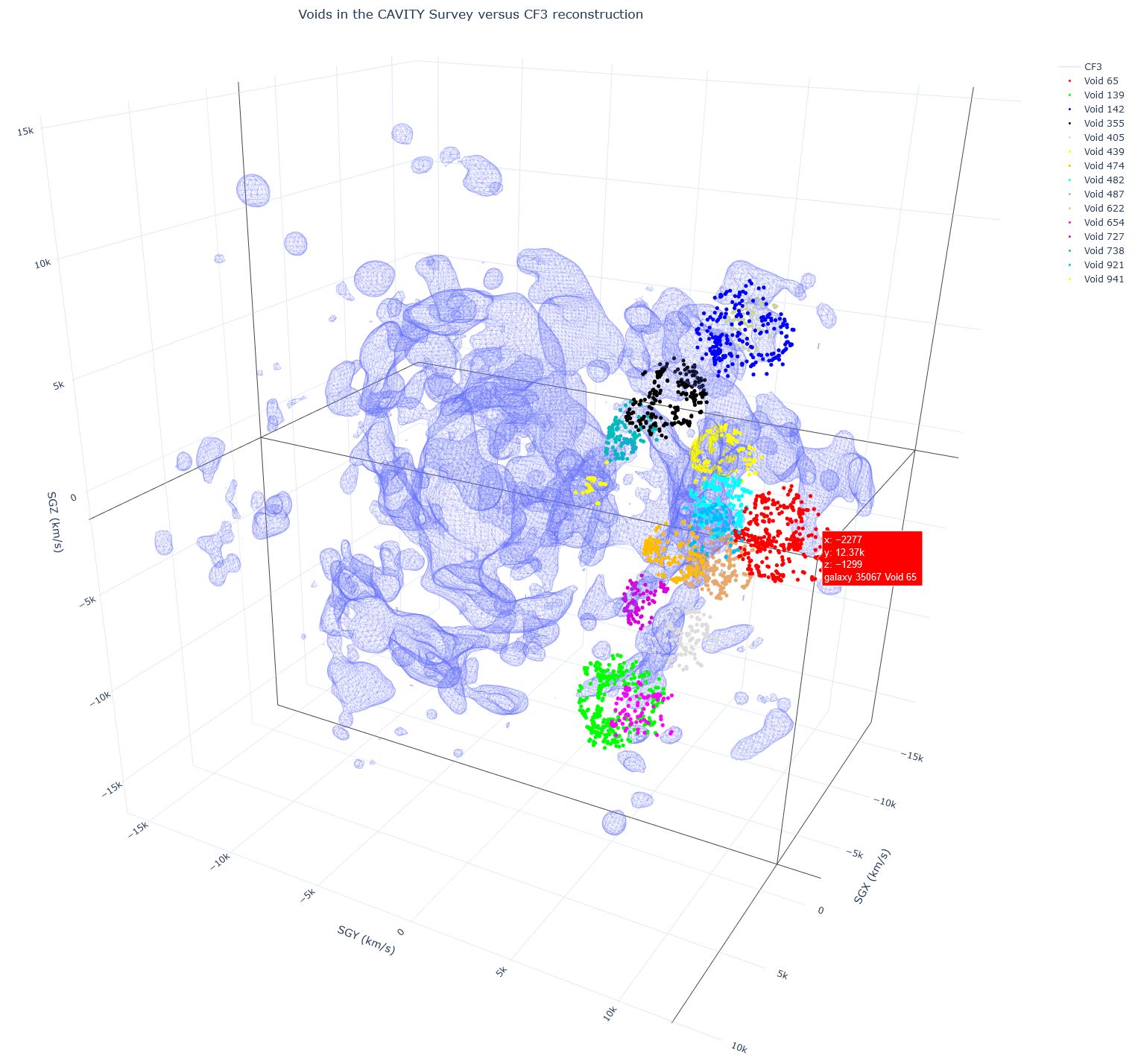}
    \caption{Interactive 3D visualization of the distribution of nearby voids \href{http://irfu.cea.fr/Projets/COAST/CAVITY-cf3.html}{[Start Interaction]}. The positions of the CAVITY survey target galaxies are given by markers colored according to their void membership as indicated in the column displayed on the right-hand side. In the online version, hovering over the galaxy markers reveals the galaxy positions, identifiers, and void memberships. The wireframe polygon is a high-density ($\delta_m =1.3$) isosurface of the reconstructed CosmicFlows-3 overdensity field. Also in the online version, the user can rotate, pan, and zoom in and out using the mouse. Single-click or double-click on the elements listed on the right-hand side column hides them or singles them out from the scene.}
    \label{fig:fig_interactive}
\end{figure*}

\begin{figure*}[ht]
    \centering
    \includegraphics[scale=0.23]{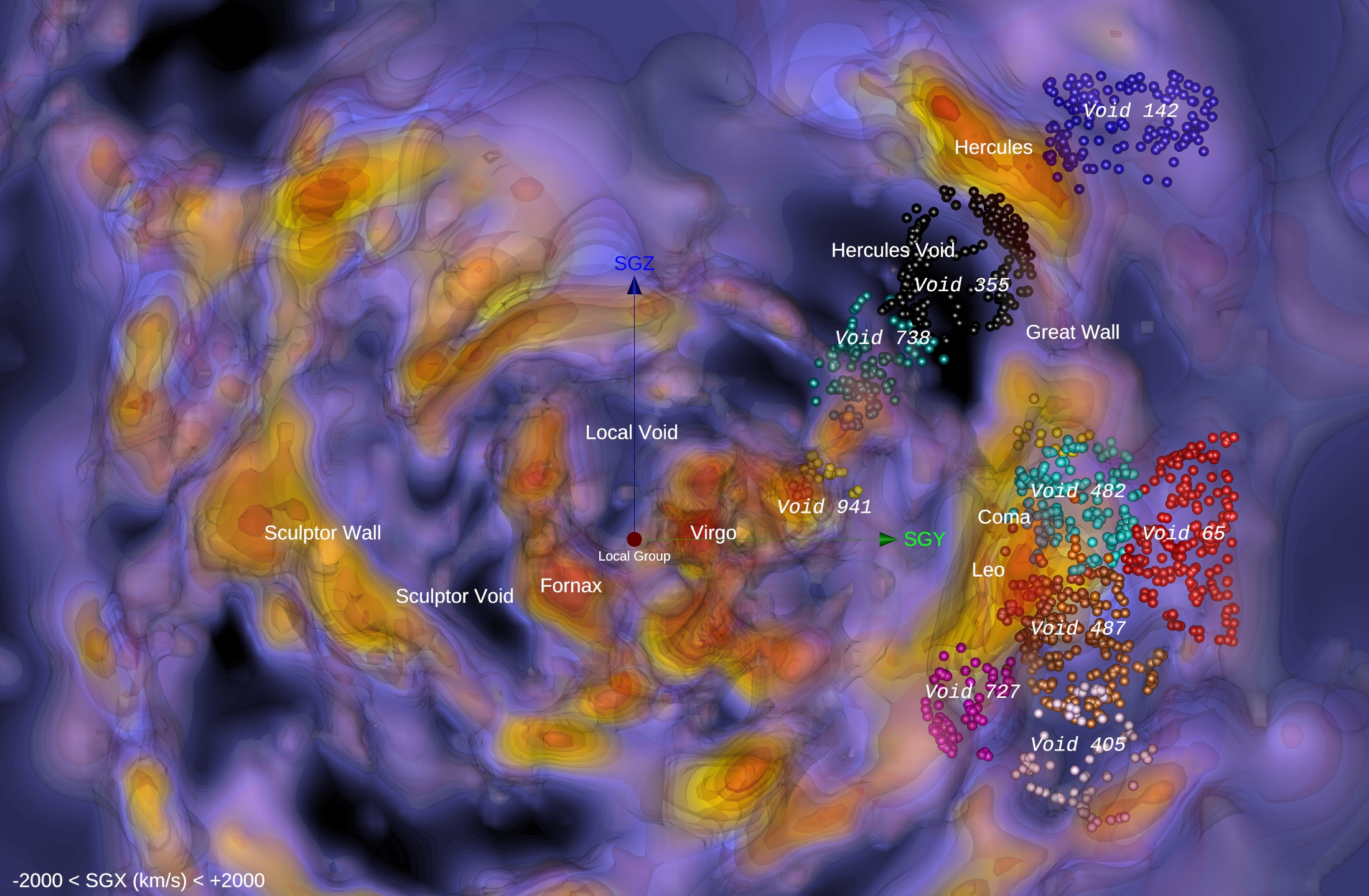}
    \caption{Map of a sample void against  the reconstructed CosmicFlows-3 density field. Map of galaxy and mass distribution within a slice $-2000 < SGX < 2000$ km/s. Galaxy markers are colored according to their void membership. Scale and orientation are given by the 5000 km/s long green (SGY) and blue (SGZ) arrows emanating from our position, associated with the cardinal axes of the Supergalactic Coordinate System.}
    \label{fig:fig_slice}
\end{figure*}

\begin{figure*}[ht]
    \centering
    \includegraphics[scale=0.23]{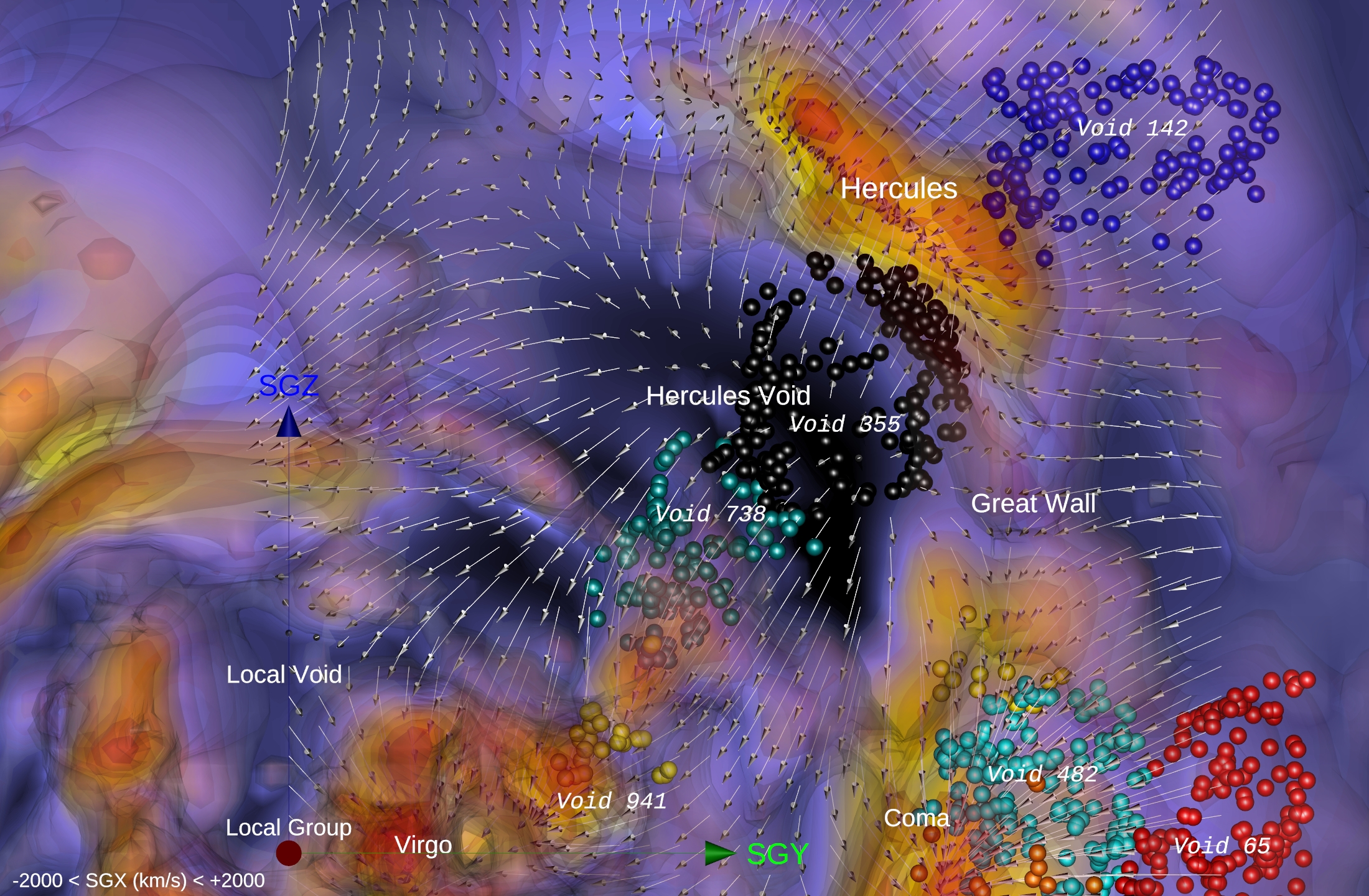}
    \caption{Focus on the Hercules region. The positions of galaxies are plotted against the reconstructed CosmicFlows-3 density contrast and velocity field, within a slice $-2000 < SGX < 2000$ km/s. Galaxy markers are given distinct colors as a function of their void membership. Scale and orientation are given by the 5000 km/s long green (SGY) and blue (SGZ) arrows emanating from our position, associated with the cardinal axes of the Supergalactic Coordinate System. The map shows that the galaxies within Void 355 and Void 738 are subject to the evacuation of matter from the Hercules Void, as mapped by the divergent flow at this location. Voids 355 and 738 belong to the same underdense region and may well 
be counted as a single "Hercules void".}
    \label{fig:fig_zoom_hercules}
\end{figure*}

Figure~\ref{fig:fig_interactive} is an interactive plot that the reader can use to better grasp the locations of the seven voids in comparison with
the matter density contrast distribution as computed using the CosmicFlows-3 catalog. Figure~\ref{fig:fig_slice} shows the supergalactic SGY--SGZ
plane orientation with a thickness of $-2000 < SGX < 2000$ km/s. This plane helps to visualize how voids 355 and 738 belong to the same underdense region and may well 
be counted as a single "Hercules void". Two voids are located near the back-side infall of Coma supercluster, namely 487 and 727. Their central
overdensity as computed from CF3 is directly linked to this large-scale structure.

\subsection{Hercules supercluster and void 355}
Figure~\ref{fig:fig_zoom_hercules} shows the region around the Hercules supercluster. Three of the voids of our sample are located around this large-scale mass concentration;  namely voids 355 and 738 on the nearer side, and void 142 on the far side of the supercluster. Of importance for our purpose are the locations of these three voids: while all are near the Hercules supercluster, they are more precisely located in the underdense interior of the surrounding mass distribution. An
exciting goal of the CosmicFlows peculiar velocity project is to fully dynamically determine the total mass of
the Hercules supercluster \citep{Dupuy2023}.

\begin{figure*}[ht]
    \centering
    \includegraphics[scale=0.5, angle=-90]{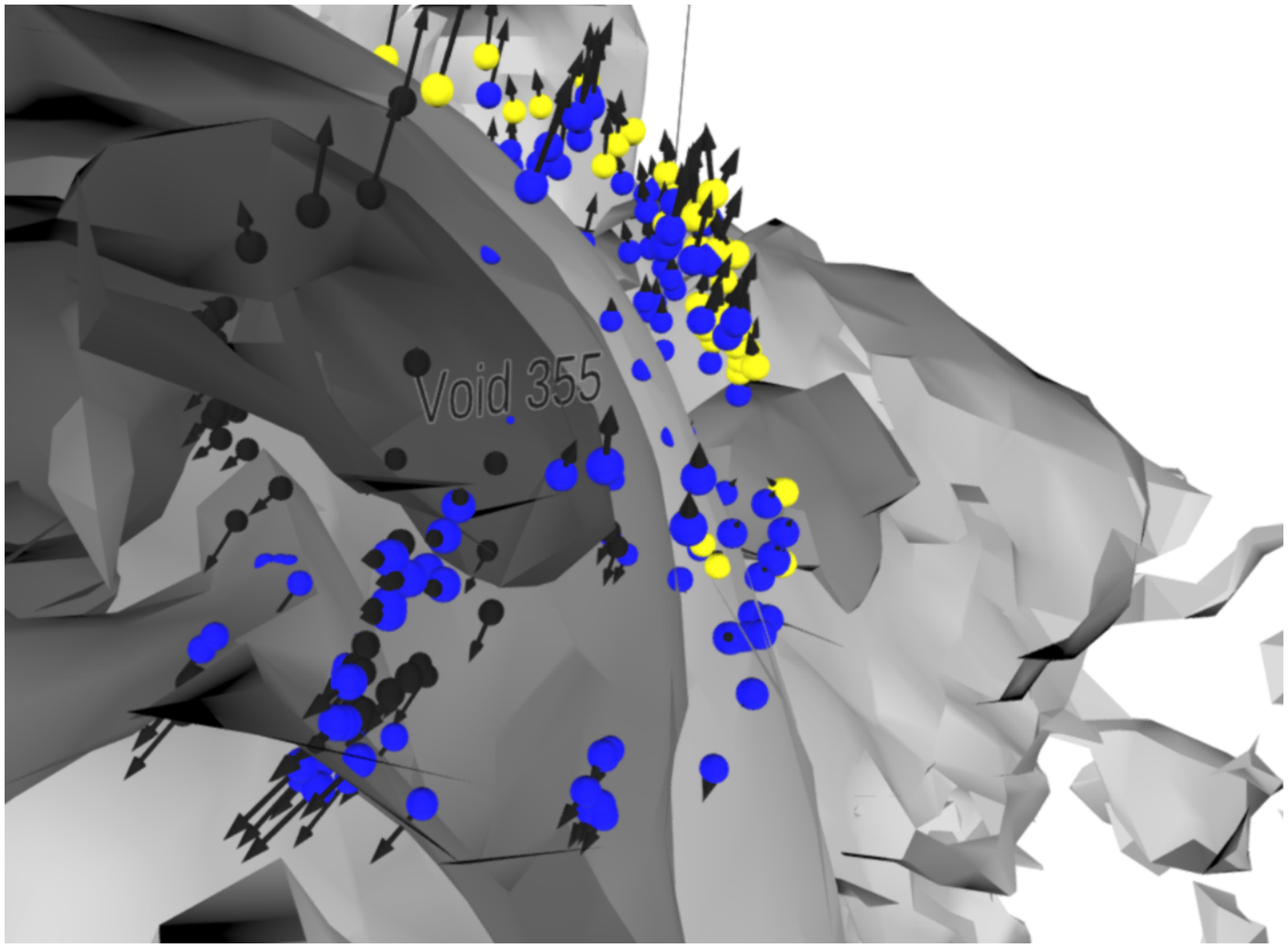}
    \caption{ V-web environment as computed using the shear tensor of the CF3 peculiar velocity field for void 355. The color code of the target galaxies is as follows: black, blue, yellow, and red identify galaxies in a V-web environment classified as empty, sheet, filament, or node, respectively. Isosurfaces in light gray to dark gray correspond to full-matter contrast levels of $\delta_m$=-0.3, -0.7, and -1.1 respectively. One can clearly see the evacuation of the local flow (the speed from the center of the vacuum was sustracted). The interactive version of the figure is available here: \href{https://sketchfab.com/3d-models/CAVITY-void-355-vs-cf3-local-flow-web-e487c0028f9d47968705aa515242b341}{[V-web environment of Void 355]}.}
    \label{fig-fig_vweb_355}
\end{figure*}

Figure~\ref{fig-fig_vweb_355} presents a zoom onto the cosmic V-web \citep{2017ApJ...845...55P} reconstruction, which was computed using the shear tensor of the
peculiar velocity field in and around void 355. The color code of the target galaxies is as follows: black, blue, yellow, and red galaxies are those that reside in a V-web environment classified as empty, sheet, filament, or node, respectively. Isosurfaces colored light gray to dark gray correspond to full matter contrast levels of $\delta_m$=-0.3, -0.7, and -1.1, respectively. We can clearly see the galaxies in black (V-Web type: "void") in the center of the void, the blue ones ("sheet") on the periphery,
and some yellow ("filament") on the side of Hercules/Great Wall, and there are no galaxies in red (classified as "knot"). The CF3 V-web reveals the very coherent
dynamical pattern of this void 355, where both galaxy and matter-density profiles are in perfect agreement (see Fig.~\ref{fig:CAVITY_densP_gridandgals_norm} and Section~\ref{sec:emptiness}). Also, Fig.~\ref{fig-fig_vweb_355} clearly reveals the evacuation by means of the pattern of velocity vectors (black). These arrows depict the local flow with respect to the center of the underdensity. The interactive version of the figure is made available at \href{https://sketchfab.com/3d-models/CAVITY-void-355-vs-cf3-local-flow-web-e487c0028f9d47968705aa515242b341}{[V-web environment of Void 355]}.

\subsection{Coma supercluster and void surroundings}
The Coma-Leo complex in the bottom right-hand corner of the mass-density map in Figure~\ref{fig:fig_slice} is surrounded by several CAVITY voids. At least three of these, voids 482, 487, and 727, are lying in or touching the overdense outskirts of the complex. We should not be surprised to find that the galaxy density and the mass density inferred from the CF3 peculiar velocity field may substantially differ (see \ref{sec:emptiness}). Below, we argue that this may be a direct manifestation of the impact of the void environment on the dynamics and evolution of voids. 

\subsection{Void sociology and hierarchical void evolution}
Visual appraisal of the void configurations in our study reveals that they provide an interesting and representative
mixture of voids in different large-scale environments. Some are more isolated or remote with respect to overdense mass concentrations (e.g., voids 405 and 738), while others reside in or near the outskirts of nearby mass concentrations. Voids 355 and 142 are close to the Hercules supercluster, and voids 482, 487, and 727 are in the outskirts of the Coma region, with 439 partly embedded inside the Coma supercluster. This means that the sample can be used to probe the effects of nearby mass concentrations on the dynamics of voids. 

Although the structure and evolution of voids is often discussed in terms of singular void configurations (see Appendix~\ref{app:appendixA}), recognizing that they are not isolated objects is of utmost importance for properly understanding them. Given the relatively mild level of their density deficit, namely $|\delta| < 1$,
external mass concentrations remain a major influence in the force inventory of voids \citep[see][]{shethwey2004,weygaert2016}. Moreover, when
we consider the population of underdensities in the mass distribution, we find that the canonical deep under-dense near-spherical void regions that
expand in each direction represent a smaller fraction of the void population. Most underdensities may expand along one or two dimensions, but contract
along the other directions \citep{shethwey2004,lavaux2010}. These voids tend to remain smaller, and may even collapse because of the surrounding over-densities.
This latter process is called {\it void-in-cloud} \citep{shethwey2004}. 

\subsubsection{Hierarchical void evolution}
On the basis of the dynamics and location, we may recognize two principal processes of void evolution \citep{shethwey2004}. {\it Void-in-void} refers to the
process whereby expanding voids merge into even larger voids, resembling the fate of  soap bubbles in a bath \citep[see e.g.,][]{dubinski1992}. For voids in
or near high-density regions that dominate their dynamical evolution, the {\it void-in-cloud} process refers to the disappearance of voids because of the
gravitational contraction and collapse induced by the environment. 

The physical context is such that there is a hierarchy of voids, with the velocity field dominated by large expanding voids, interior to which
are found smaller, often elongated voids, in particular near the edges of the large expanding voids (see Fig. 11 of \cite{weygaert2016}). These smaller voids may
contract along one or two dimensions, or even fully collapse. Also important is the fact that many of these voids are not spherical at all, but
become substantially deformed by dominant tidal influences of the surrounding mass concentrations. This process may even involve fully collapsing voids entirely embedded in overdense structures. 

Hierarchical void evolution leads to a situation in which it may not be straightforward to relate velocity flow and density in and around voids, as much of the (filtered) flow field includes the dynamical influence of the nearby high-density regions. A detailed study and analysis of
the corresponding properties of the hierarchical embedding of void flows, particular in terms of the velocity divergence, is presented by \cite{aragon2013}  and also see \cite{aragon2010}. 

\subsection{Void environment and dynamical impact}
Following the observation of diverse void environments in our sample, and the implications for their dynamical evolution and fate, a quantitative and statistical analysis of the environmental imprint is presented in the following section. The results show that the visual inspection of the position of voids compared to the well-known large-scale structures in the local Universe is indeed confirmed quantitatively. The identification of voids in
galaxy-redshift surveys appears to lead to the inclusion of voids that partake in the {\it void-in-cloud} process (see sect. 5), that is, voids that are contracting ---along one, two, or three directions--- because of the over-dense surroundings. In simulations, these are usually the small voids near the filamentary and wall-like boundaries of large voids \citep{shethwey2004}.  Early indications of this were found
in the SDSS survey by the redshift space correlation function analysis by \cite{paz2013}. In other words, our analysis indicates that this is the state
of some of the CAVITY voids. 

\section{Emptiness of voids: Results} 
\label{sec:emptiness}
In this section, we compare the $\delta_g$ values from galaxy redshift counts to the CF3 reconstruction $\delta_m$ contrast field on the basis of the galaxy and mass-density profiles of the seven voids included in our sample. Figure~\ref{fig:CAVITY_densP_gridandgals_norm} shows the galaxy counts around the seven voids (left panels) together with the mass and galaxy number density profiles (right panels). 

The left panels of fig.~\ref{fig:CAVITY_densP_gridandgals_norm} show the number of galaxies in the galaxy samples ---namely SDSS, SDSS void galaxies only,
and CosmicFlows third and fourth editions--- as a function of radial distance (in normalized units) $r/R_v$. For six voids, we see the expected pattern of a steeply increasing number of galaxies as a function of radius around a near-empty void interior. Only void 941, which is found near the outskirts of the Virgo cluster (see Fig.~\ref{fig:fig_slice}), displays a slightly different behavior. Many galaxies are found within half its effective radius $R_v$.

The galaxy number-density profiles (purple dashed lines) and mass-density profiles (solid blue lines) reveal a different story. The profiles in all seven right-hand panels agree on their tendency towards the global mean density value of $\delta_m$=0 at $r/R_v > 3$ (the average of the density field over all dimensions in the full CF3 grid gives a value of a mean $\delta_m = 0.005$). This implies that all void configurations consist of a central void surrounded by overdense structures within $1<r/R_v<3$. Most importantly, in at least half of the sample of voids, there is a strong difference between the computed galaxy number-density profile and the mass-density profile inferred from the peculiar velocity field.

While the galaxy number-density profiles for almost all the voids of our sample display the well-known ``bucket-shaped'' inner profile, the CF3-implied mass profiles show a different behavior. For half of our small sample of seven voids, there is strong disagreement in the void emptiness when computed from galaxy counts (empty) and from peculiar velocity dynamics (overdensity near the center). Only in the case of void 355 do we find perfect agreement between galaxy number-density and mass-density profiles. This may relate to the fact that void 355 is a relatively large and well-defined void of which a significant part is located near the center of the Hercules void-mass underdensity. On the other hand, in the case of three or even four voids, we find implied mass overdensities near the center. In summary, the blue solid lines in
fig.~\ref{fig:CAVITY_densP_gridandgals_norm} show that the CF3-reconstructed matter profiles (computed using equation~\ref{delta_CF3})
do not systematically display underdense regions near the void centers. As discussed in section~\ref{sec:environment}, some of these voids have centers
located in overdense regions in CF3.

  When analyzing and interpreting these results, it is important to take into account the differences between the probes used to trace
  the CF3 and the CAVITY probes. The detected voids in different galaxy samples are sensitively dependent on the number density and nature of the
  galaxy population in those samples \citep[see]{peebles2001,gottloeber2003,tinker2006}. It is self-evident that voids in more diluted galaxy
  samples will be
  larger  on average, as these samples lack the spatial resolution to resolve the smaller voids. A more profound influence is the fact that the spatial
  clustering of galaxies is also sensitively dependent on the galaxies in the sample. Brighter and heavier galaxies are more strongly clustered,
  and there will therefore be larger  "cavities" in their spatial distribution.  A recent study found that the void population in different galaxy
  populations is indeed dependent on higher-order clustering properties of the galaxy population, in excess of its two-point clustering
  properties \citep{wilding2022}. These authors found proof of a strongly systematic dependence of the void population on the topological characteristics of the spatial
  galaxy distribution. In other words, voids detected in different galaxy populations are strongly affected by a
  {topological bias}.

  Evidence that this subtle galaxy bias affects the void population and inferred void-density profiles was pointed out in the
  analysis of voids in the SDSS galaxy survey by \cite{ricciardelli2013}. These authors found that differences in void probes lead to systematic
  differences between void-density profiles based on galaxy counts and voids extracted from the density distribution in simulations. This
  may certainly be a factor of relevance in the comparison between the CAVITY and CF3 voids in the present study. However, while this bias
can undoubtedly play a role, it cannot explain our results since the levels of
  anti-bias required would be unrealistic \citep{braun1988}.

\section{Discussion and conclusions  }
In the presented analysis, we compare the galaxy distribution in and around a small sample of seven voids from the CAVITY void galaxy survey with the dynamically inferred mass distribution in and around voids. The latter distribution is based on the mass reconstruction from the velocity flow field measured using the Cosmicflows-3 catalog. The comparison between the mass and galaxy distributions around these voids, in conjunction with the map of the large-scale mass distribution and features in and around these voids in the Local Universe, enables us to assess the impact of environment on the dynamics and hierarchical evolution of the void population. The results of this comparison demonstrate the reality and importance of the {void-in-cloud} process in the buildup of the web-like matter distribution in the Universe.
Our findings also reveal the importance of distinguishing these voids from the dominant fully expanding voids ---the result of the {void-in-void} merging process---
when we seek to study the role of voids in a cosmological setting \citep[e.g.,][]{weygaert2016,pisani2020} or when assessing their influence on galaxy formation and the galaxy star formation history \citep[see][for theoretical treatments]{goldberg2004,lackner2012}.

The sample of seven CAVITY voids was identified with a classical void finder from the SDSS redshift survey. Using a similar algorithm we confirm
that local voids are empty of galaxies near their center and roughly up to their effective radius. However, a different picture emerges when studying the
velocity field in and around these voids. When assessing the dynamics of the void regions, and computing the matter content from the measured CF3 velocity flows, in half of the cases, we find that on the corresponding scale of the velocity flows these void regions are not underdense. 

There are several reasons why the mass and galaxy distribution around the sample voids may be different.
One factor is that of galaxy bias, with the galaxy population not entirely reflecting the underlying mass distribution. While we do not exclude this factor, we find that the strong levels of anti-bias that would be needed to explain our results are unrealistic \citep{peebles2001,gottloeber2003,tinker2006,ricciardelli2013,wilding2022}. We believe that the overriding reason for the difference is to be found in the location of the voids with respect to their surroundings. Several voids of the sample are located in or are identified with large underdensities in the mass distribution. The most interesting ones, slightly more than half of the sample, are found at the outer regions of the Coma-Leo and Virgo mass supercluster complexes. 

A major part of the explanation for the difference between the galaxy distribution in and around these voids and the inferred large-scale mass distribution is the dynamical impact of the void environment. The void regions affiliated with large underdensities in the CF3 map of the Local Universe partake in the overall expansion of the region. The divergent velocity flow translates into a corresponding void mass-density profile. This latter is a typical manifestation of the so-called
{void-in-void} configuration that goes along with the hierarchical buildup of the void population \citep{shethwey2004}.

However, in our void sample, we find that the majority of CAVITY voids most likely belong to the class of voids that are not expanding in three dimensions, and may even contract along one or two dimensions. These {void-in-cloud} voids \citep{shethwey2004} represent the majority of the underdense regions in the mass distribution, and in the hierarchical buildup of structure are often found at the boundaries of large voids and surrounding overdense filaments and walls. The flow in and around these voids is largely dominated by the dynamical ---tidal--- influence of the nearby overdense filaments and walls. This translates into an anisotropic and in some cases even fully collapsing flow field in and around the void, which explains the difference between the galaxy underdensity and the implied mass distribution, which includes the contribution from the surrounding overdense large-scale structure. Hence, while on the large linear scales we find the overdensity as it takes into account the large-scale surroundings, on smaller scales we would recover the underdensity that is also seen in the galaxy distribution. This explains why the galaxy counts may hint at a local cavity, while the CF3 velocity flow would suggest otherwise.

Within the context of the hierarchically evolving void population, and the role of the surroundings in the evolution of voids, we may also find the merging of voids with overdense walls or filaments. A recent theoretical study \citep{2021ApJ...920L...2V} found that $\sim  10\%$ of the mass in voids at $z=0$ may be
accreted from overdense regions, with this value even reaching beyond $35\%$ for a significant fraction of voids.

\medskip
While the present analysis of a limited void sample demonstrates the potential for studying voids in relation to their large-scale environment, we expect
exciting and statistically representative results for the Local Universe void population from the recently released CosmicFlows-4 dataset of galaxy distances, which 
includes the entire SDSS volume \citep{2023A&A...670L..15C}. 

\begin{figure*}[htp]
\includegraphics[width=0.9\textwidth]{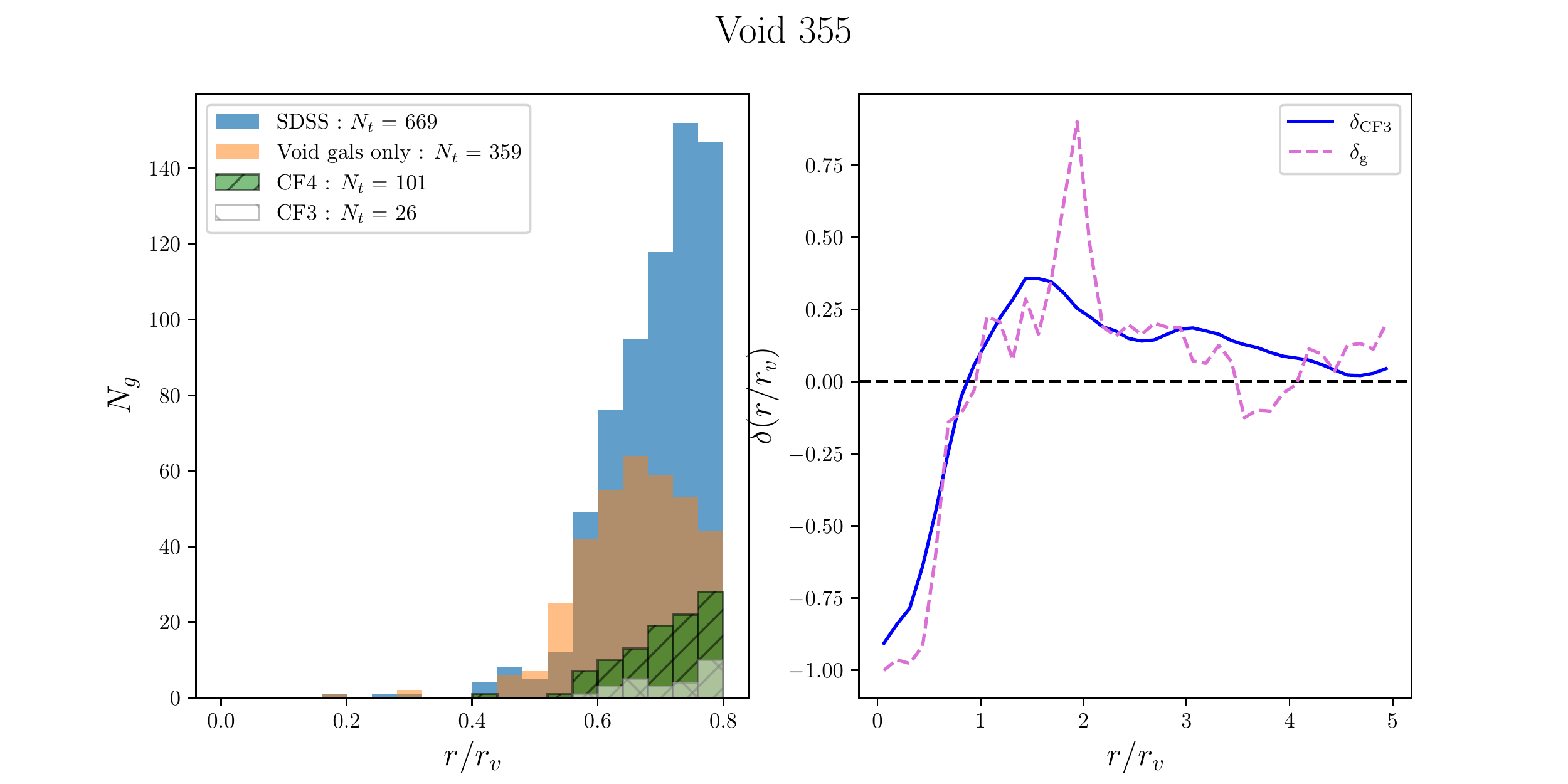} 

\includegraphics[width=0.9\textwidth]{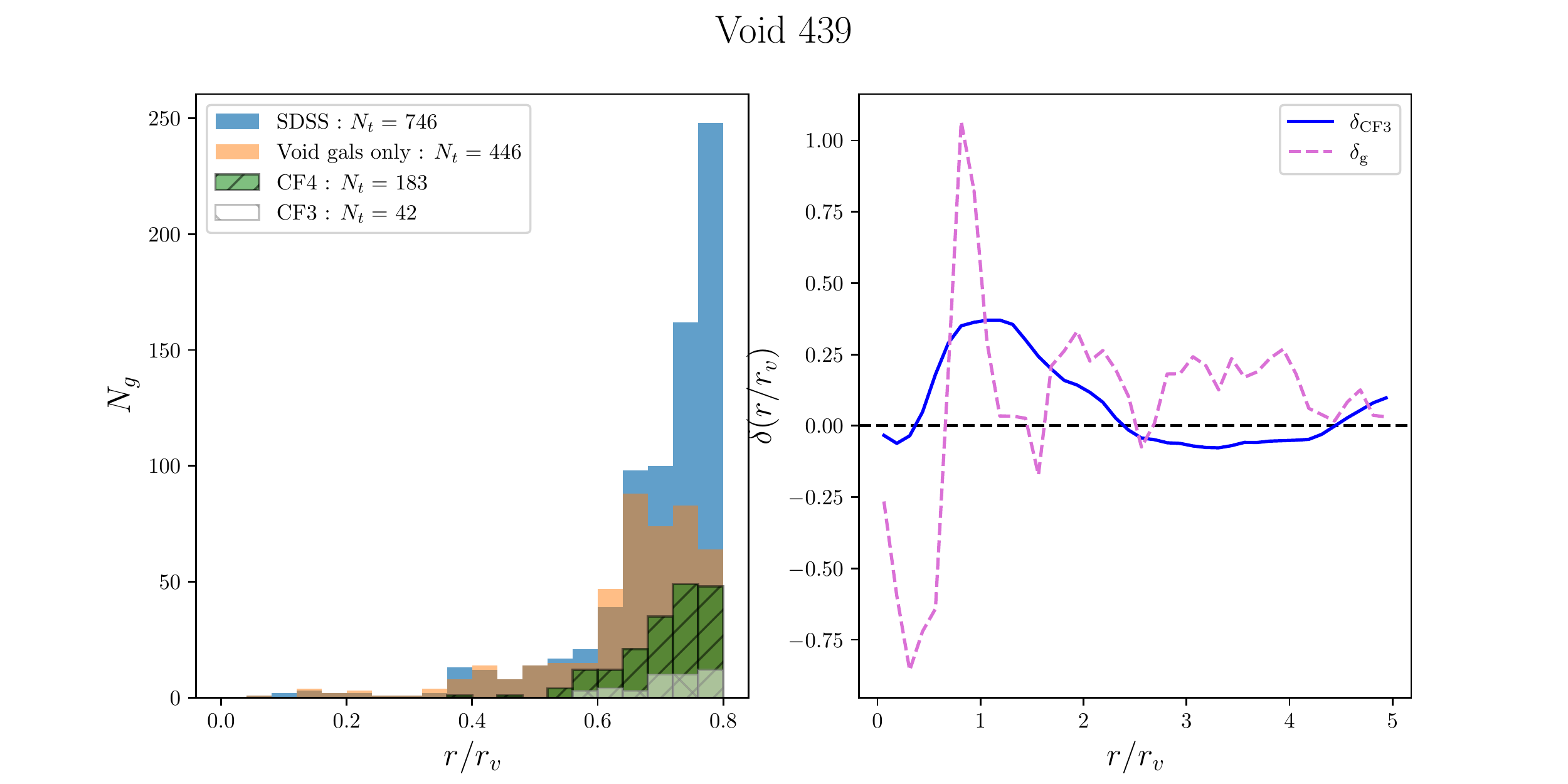}

\includegraphics[width=0.9\textwidth]{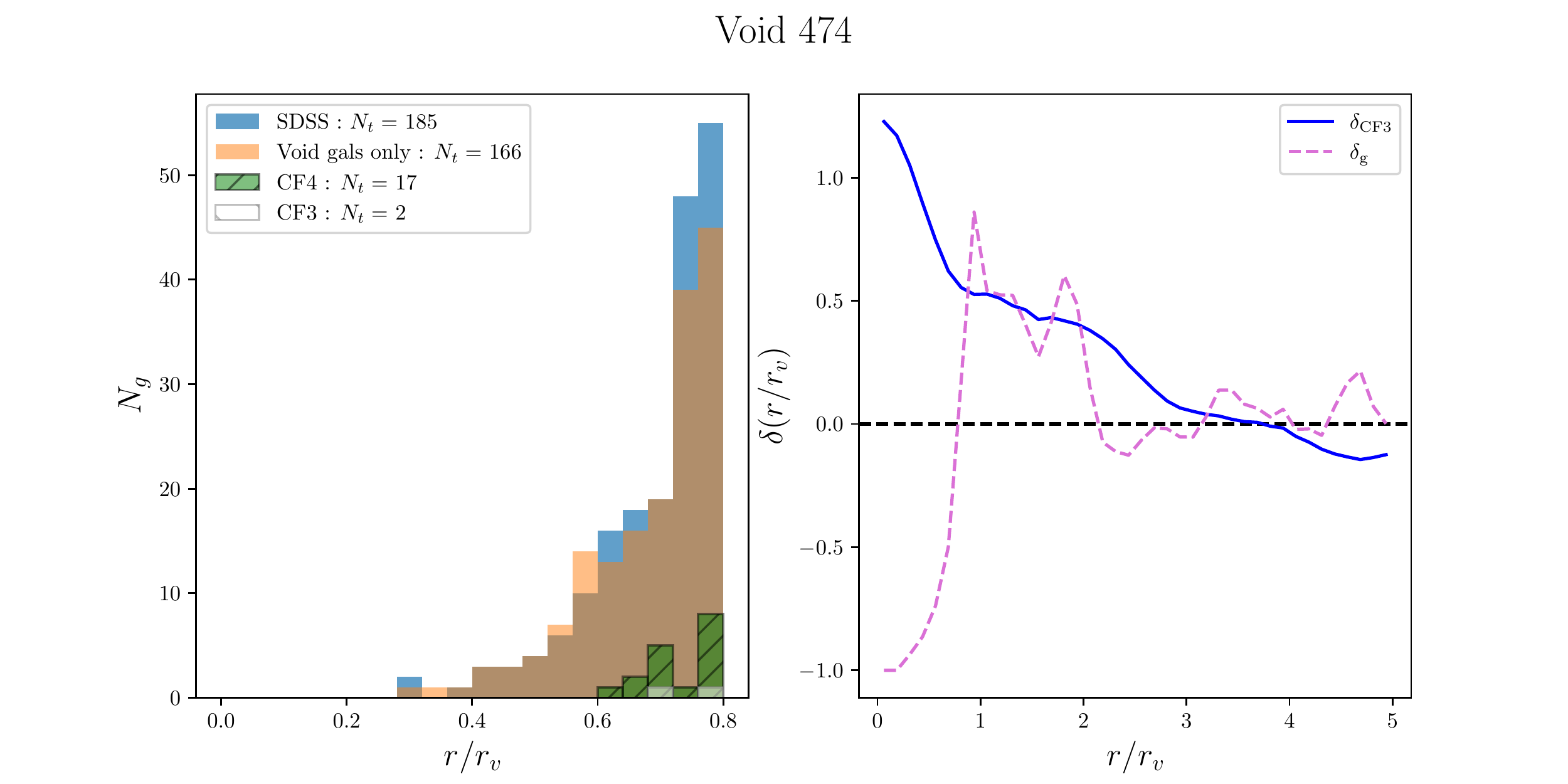} 
\end{figure*}

\begin{figure*}[htp]\ContinuedFloat
\includegraphics[width=0.9\textwidth]{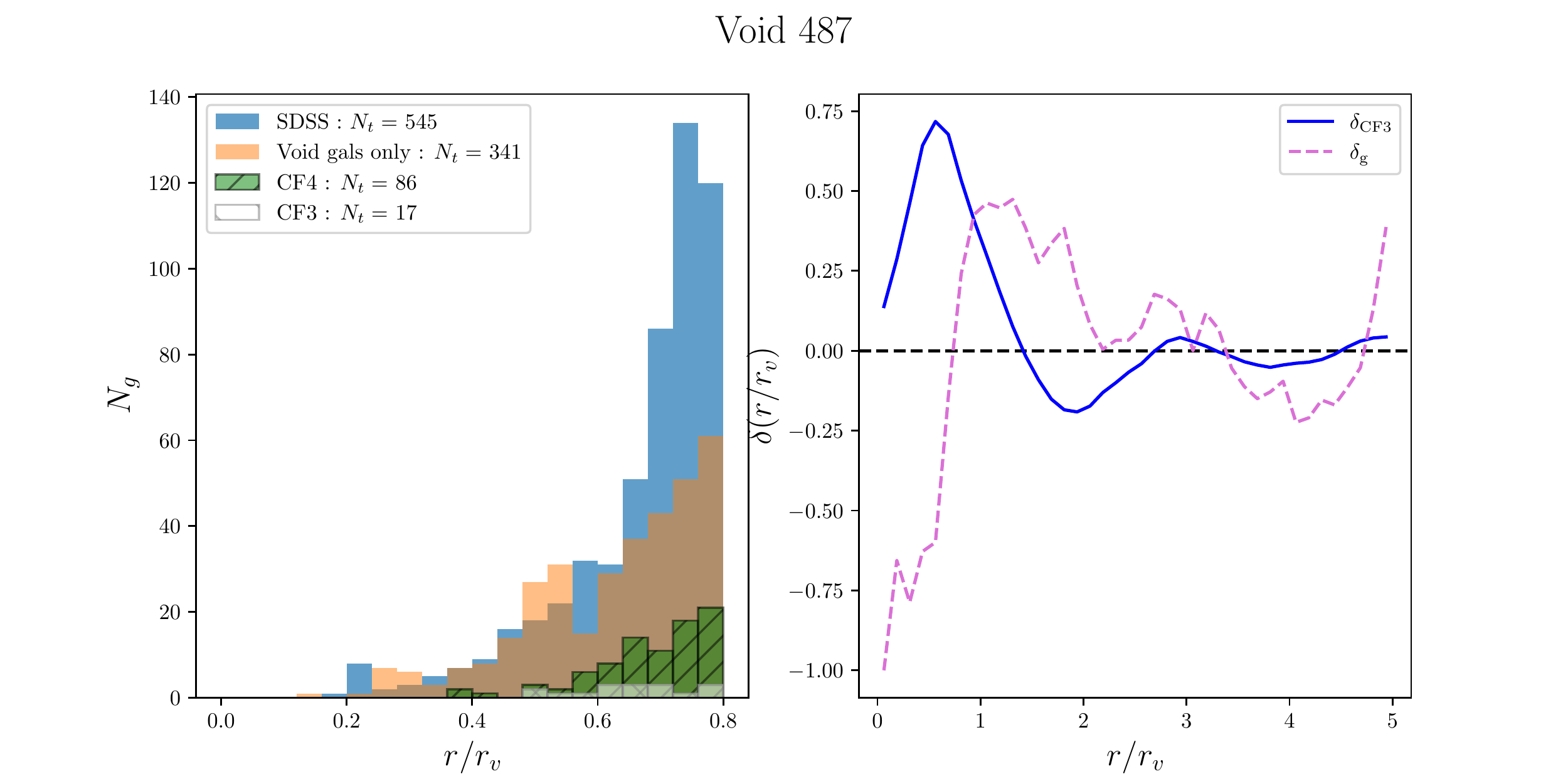}

\includegraphics[width=0.9\textwidth]{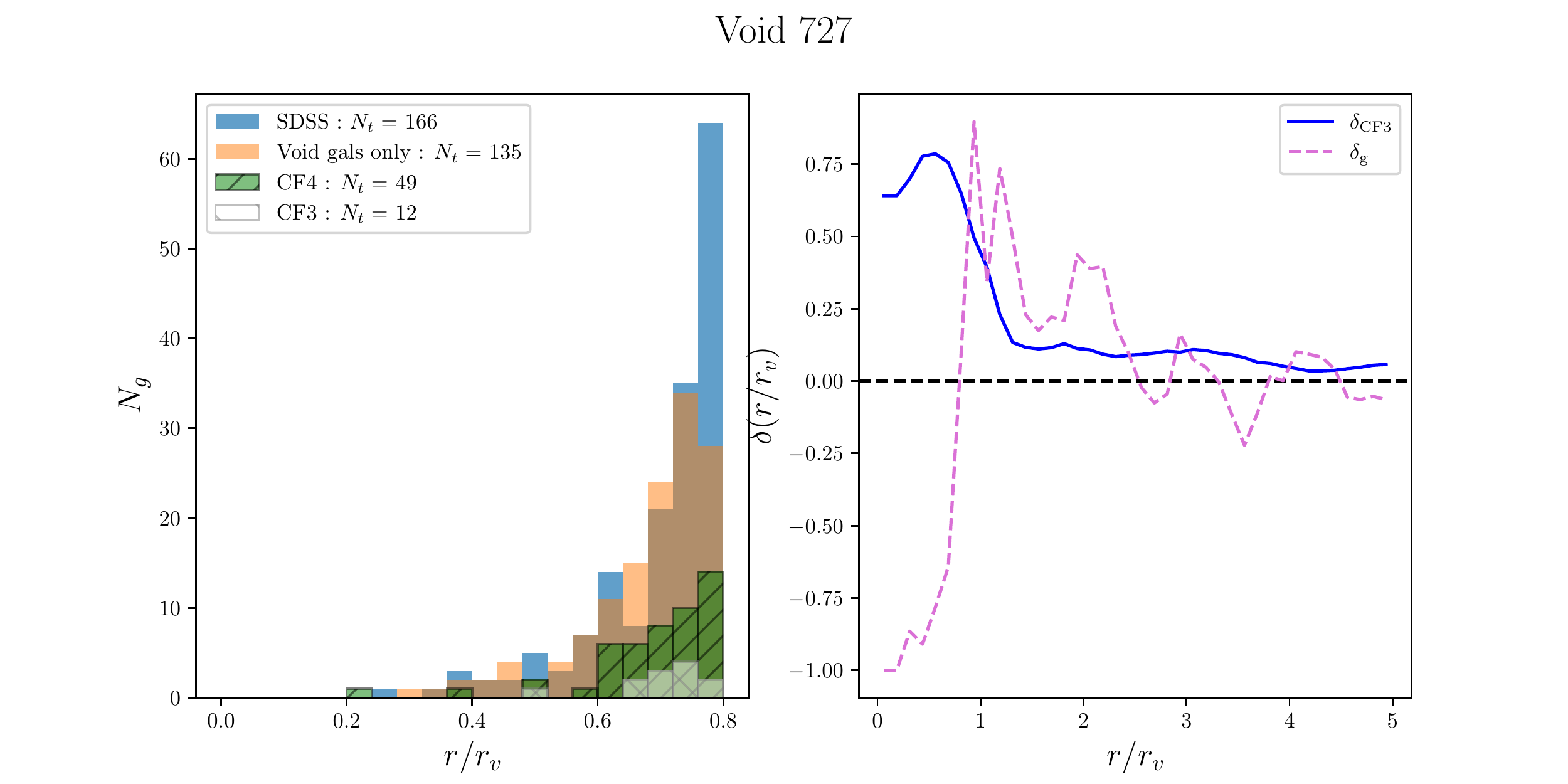} 

\includegraphics[width=0.9\textwidth]{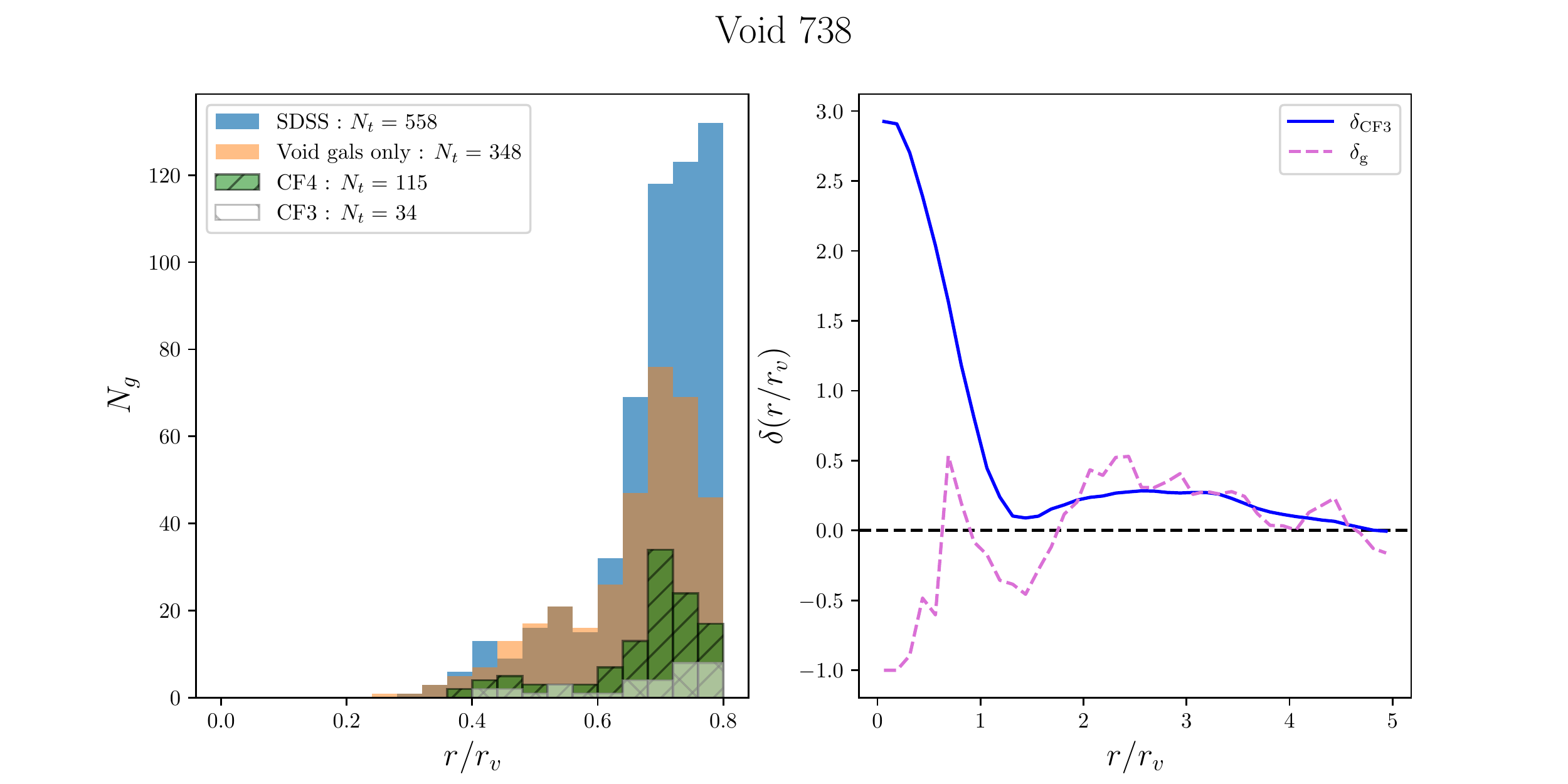}
\end{figure*}
\begin{figure*}[htp]\ContinuedFloat
\includegraphics[width=0.9\textwidth]{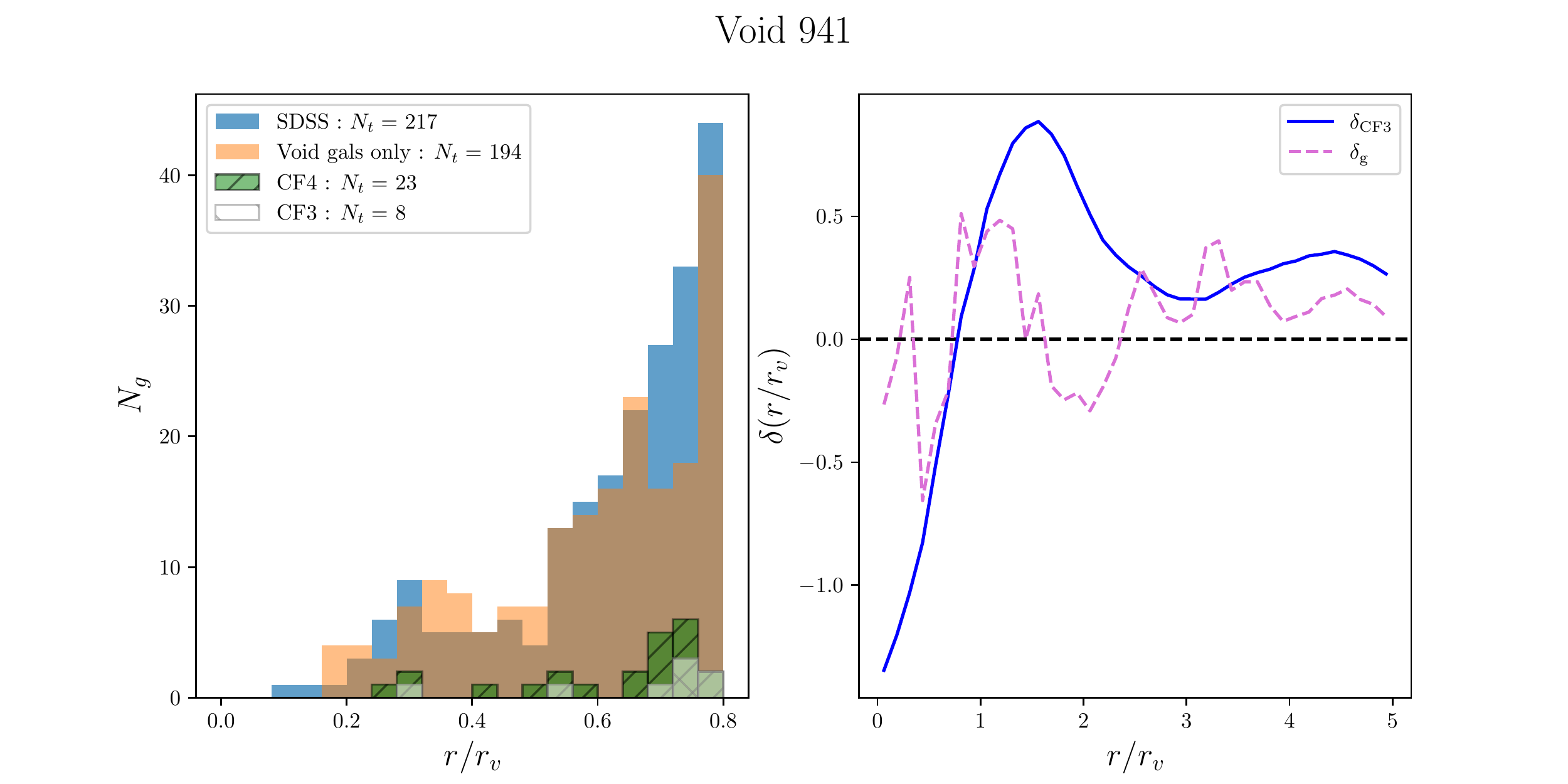}
 \caption{Void radial density profiles. For each of the seven SDSS/CAVITY nearby voids that are included in the CosmicFlows-3 volume, we show the radial number of galaxies  in the
left panel and the matter content in the right panel computed from CF3 (blue) and from the galaxy number density in SDSS (pink). All seven voids are empty of galaxies near their center and roughly up to their effective radius. However, four voids (474, 487, 727 and 738) display CF3-computed overdensities of matter in their center.}
 \label{fig:CAVITY_densP_gridandgals_norm}
\end{figure*}

\begin{acknowledgements}
HC is grateful to the Institut Universitaire de France for its support. HC, MA, DG acknowledge support from the CNES. 

RvdW is grateful to Miguel Aragon-Calvo, Job Feldbrugge, Roi Kugul, Bernard Jones, Georg Wilding and Raul Bermejo on numerous discussions on the nature of voids and the role of topological bias. 

E.F. 
is grateful to the Spanish 'Ministerio de Ciencia e Innovaci\'on and from the European Regional Development Fund (FEDER) via grants PID2020-224414GB-I00 and PID2020-113689GB-I00, and from the 'Junta de Andaluc\'{i}a' (Spain) local government through the FQM108 and A-FQM-510-UGR20 projects.

L.G. acknowledges financial support from the Spanish Ministerio de Ciencia e Innovaci\'on (MCIN), the Agencia Estatal de Investigaci\'on (AEI) 10.13039/501100011033, and the European Social Fund (ESF) "Investing in your future" under the 2019 Ram\'on y Cajal program RYC2019-027683-I and the PID2020-115253GA-I00 HOSTFLOWS project, from Centro Superior de Investigaciones Cient\'ificas (CSIC) under the PIE project 20215AT016, and the program Unidad de Excelencia Mar\'ia de Maeztu CEX2020-001058-M.

RGB acknowledges financial support from the grants CEX2021-001131-S funded by MCIN/AEI/10.13039/501100011033 and PID2019-109067-GB100.

KK gratefully acknowledges funding from the German Research Foundation (DFG) in the form of an Emmy Noether Research Group (grant number KR4598/2-1, PI Kreckel).

SP and VQ acknowledge support by the Agencia Estatal de Investigaci\'on Espa\~{n}ola (AEI; grant PID2019-107427GB-C33), by the Ministerio de Ciencia e Innovaci\'on (MCIN) within the Plan de Recuperaci\'on, Transformaci\'on y Resiliencia del Gobierno de Espa\~{n}a through the project ASFAE/2022/001, with funding from European Union NextGenerationEU (PRTR-C17.I1), and by the Generalitat Valenciana (grant PROMETEO/2019/071).

JR acknowledges support from the State Research Agency (AEI-MCINN) of the Spanish Ministry of Science and Innovation under the grant "The structure and evolution of galaxies and their central regions" with reference PID2019-105602GBI00/10.13039/501100011033, funding granted for his Margarita Salas Fellowship by the Ministry of Universities granted by Order UNI/551/2021 of May 26, as well as funding by the European Union-Next Generation EU Funds.

MSP acknowledges support from the Spanish Ministry of Science and Innovation through project AYA2017–88007–C3–2–P.

\end{acknowledgements}

\bibliography{Cavity}

\begin{thebibliography}{69}
\expandafter\ifx\csname natexlab\endcsname\relax\def\natexlab#1{#1}\fi

\bibitem[{{Aragon-Calvo} \& {Szalay}(2013)}]{aragon2013}
{Aragon-Calvo}, M.~A. \& {Szalay}, A.~S. 2013, \mnras, 428, 3409

\bibitem[{{Aragon-Calvo} {et~al.}(2010){Aragon-Calvo}, {van de Weygaert},
  {Araya-Melo}, {Platen}, \& {Szalay}}]{aragon2010}
{Aragon-Calvo}, M.~A., {van de Weygaert}, R., {Araya-Melo}, P.~A., {Platen},
  E., \& {Szalay}, A.~S. 2010, \mnras, 404, L89

\bibitem[{{Bermejo} {et~al.}(2022){Bermejo}, {Wilding}, {van de Weygaert},
  {Jones}, {Vegter}, \& {Efstathiou}}]{wilding2022}
{Bermejo}, R., {Wilding}, G., {van de Weygaert}, R., {et~al.} 2022, arXiv
  e-prints, arXiv:2206.14655

\bibitem[{{Bertschinger}(1985)}]{bertschinger1985}
{Bertschinger}, E. 1985, \apj, 295, 1

\bibitem[{{Beygu} {et~al.}(2016){Beygu}, {Kreckel}, {van der Hulst}, {Jarrett},
  {Peletier}, {van de Weygaert}, {van Gorkom}, \& {Aragon-Calvo}}]{beygu2016}
{Beygu}, B., {Kreckel}, K., {van der Hulst}, J.~M., {et~al.} 2016, \mnras, 458,
  394

\bibitem[{{Beygu} {et~al.}(2017){Beygu}, {Peletier}, {van der Hulst},
  {Jarrett}, {Kreckel}, {van de Weygaert}, {van Gorkom}, \&
  {Aragon-Calvo}}]{beygu2017}
{Beygu}, B., {Peletier}, R.~F., {van der Hulst}, J.~M., {et~al.} 2017, \mnras,
  464, 666

\bibitem[{{Blumenthal} {et~al.}(1992){Blumenthal}, {da Costa}, {Goldwirth},
  {Lecar}, \& {Piran}}]{blumenthal1992}
{Blumenthal}, G.~R., {da Costa}, L.~N., {Goldwirth}, D.~S., {Lecar}, M., \&
  {Piran}, T. 1992, \apj, 388, 234

\bibitem[{{Bond} {et~al.}(1996){Bond}, {Kofman}, \& {Pogosyan}}]{bond1996}
{Bond}, J., {Kofman}, L., \& {Pogosyan}, D. 1996, \nat, 380, 603

\bibitem[{{Borzyszkowski} {et~al.}(2017){Borzyszkowski}, {Porciani},
  {Romano-D{\'\i}az}, \& {Garaldi}}]{borz2017}
{Borzyszkowski}, M., {Porciani}, C., {Romano-D{\'\i}az}, E., \& {Garaldi}, E.
  2017, \mnras, 469, 594

\bibitem[{{Bos} {et~al.}(2012){Bos}, {van de Weygaert}, {Dolag}, \&
  {Pettorino}}]{bos2012}
{Bos}, E.~G.~P., {van de Weygaert}, R., {Dolag}, K., \& {Pettorino}, V. 2012,
  \mnras, 426, 440

\bibitem[{{Braun} {et~al.}(1988){Braun}, {Dekel}, \& {Shapiro}}]{braun1988}
{Braun}, E., {Dekel}, A., \& {Shapiro}, P.~R. 1988, \apj, 328, 34

\bibitem[{{Cai} {et~al.}(2015){Cai}, {Padilla}, \& {Li}}]{cai2015}
{Cai}, Y.-C., {Padilla}, N., \& {Li}, B. 2015, \mnras, 451, 1036

\bibitem[{{Cautun} {et~al.}(2016){Cautun}, {Cai}, \& {Frenk}}]{cautun2016}
{Cautun}, M., {Cai}, Y.-C., \& {Frenk}, C.~S. 2016, \mnras, 457, 2540

\bibitem[{{Cautun} {et~al.}(2014){Cautun}, {van de Weygaert}, {Jones}, \&
  {Frenk}}]{cautun2014}
{Cautun}, M., {van de Weygaert}, R., {Jones}, B. J.~T., \& {Frenk}, C.~S. 2014,
  \mnras, 441, 2923

\bibitem[{{Courtois} {et~al.}(2023){Courtois}, {Dupuy}, {Guinet}, {Baulieu},
  {Ruppin}, \& {Brenas}}]{2023A&A...670L..15C}
{Courtois}, H.~M., {Dupuy}, A., {Guinet}, D., {et~al.} 2023, \aap, 670, L15

\bibitem[{{Courtois} {et~al.}(2012){Courtois}, {Hoffman}, {Tully}, \&
  {Gottl{\"o}ber}}]{courtois2012}
{Courtois}, H.~M., {Hoffman}, Y., {Tully}, R.~B., \& {Gottl{\"o}ber}, S. 2012,
  \apj, 744, 43

\bibitem[{{Davis} \& {Peebles}(1983)}]{davispeebles1983}
{Davis}, M. \& {Peebles}, P.~J.~E. 1983, \apj, 267, 465

\bibitem[{{de Lapparent} {et~al.}(1986){de Lapparent}, {Geller}, \&
  {Huchra}}]{lapparent1986}
{de Lapparent}, V., {Geller}, M.~J., \& {Huchra}, J.~P. 1986, \apjl, 302, L1

\bibitem[{{Dom{\'\i}nguez-G{\'o}mez} {et~al.}(2022){Dom{\'\i}nguez-G{\'o}mez},
  {Lisenfeld}, {P{\'e}rez}, {L{\'o}pez-S{\'a}nchez}, {Duarte Puertas},
  {Falc{\'o}n-Barroso}, {Kreckel}, {Peletier}, {Ruiz-Lara}, {van de Weygaert},
  {van der Hulst}, \& {Verley}}]{dominguezgomez2022}
{Dom{\'\i}nguez-G{\'o}mez}, J., {Lisenfeld}, U., {P{\'e}rez}, I., {et~al.}
  2022, \aap, 658, A124

\bibitem[{{Dubinski} {et~al.}(1993){Dubinski}, {da Costa}, {Goldwirth},
  {Lecar}, \& {Piran}}]{dubinski1992}
{Dubinski}, J., {da Costa}, L.~N., {Goldwirth}, D.~S., {Lecar}, M., \& {Piran},
  T. 1993, \apj, 410, 458

\bibitem[{{El-Ad} \& {Piran}(1997)}]{el-ad_voids_1997}
{El-Ad}, H. \& {Piran}, T. 1997, \apj, 491, 421

\bibitem[{{Ganeshaiah Veena} {et~al.}(2019){Ganeshaiah Veena}, {Cautun},
  {Tempel}, {van de Weygaert}, \& {Frenk}}]{punya2019}
{Ganeshaiah Veena}, P., {Cautun}, M., {Tempel}, E., {van de Weygaert}, R., \&
  {Frenk}, C.~S. 2019, \mnras, 487, 1607

\bibitem[{{Goh} {et~al.}(2019){Goh}, {Primack}, {Lee}, {Aragon-Calvo},
  {Hellinger}, {Behroozi}, {Rodriguez-Puebla}, {Eckholm}, \&
  {Johnston}}]{goh2019}
{Goh}, T., {Primack}, J., {Lee}, C.~T., {et~al.} 2019, \mnras, 483, 2101

\bibitem[{{Goldberg} \& {Vogeley}(2004)}]{goldberg2004}
{Goldberg}, D.~M. \& {Vogeley}, M.~S. 2004, \apj, 605, 1

\bibitem[{{Gottl{\"o}ber} {et~al.}(2003){Gottl{\"o}ber}, {{\L}okas}, {Klypin},
  \& {Hoffman}}]{gottloeber2003}
{Gottl{\"o}ber}, S., {{\L}okas}, E.~L., {Klypin}, A., \& {Hoffman}, Y. 2003,
  \mnras, 344, 715

\bibitem[{{Graziani} {et~al.}(2019){Graziani}, {Courtois}, {Lavaux}, {Hoffman},
  {Tully}, {Copin}, \& {Pomar{\`e}de}}]{2019MNRAS.488.5438G}
{Graziani}, R., {Courtois}, H.~M., {Lavaux}, G., {et~al.} 2019, \mnras, 488,
  5438

\bibitem[{{Hahn} {et~al.}(2007){Hahn}, {Porciani}, {Carollo}, \&
  {Dekel}}]{hahn2007}
{Hahn}, O., {Porciani}, C., {Carollo}, C.~M., \& {Dekel}, A. 2007, \mnras, 375,
  489

\bibitem[{{Hamaus} {et~al.}(2016){Hamaus}, {Pisani}, {Sutter}, {Lavaux},
  {Escoffier}, {Wandelt}, \& {Weller}}]{hamaus2016}
{Hamaus}, N., {Pisani}, A., {Sutter}, P.~M., {et~al.} 2016, \prl, 117, 091302

\bibitem[{{Hamaus} {et~al.}(2014){Hamaus}, {Sutter}, \& {Wandelt}}]{hamaus2014}
{Hamaus}, N., {Sutter}, P.~M., \& {Wandelt}, B.~D. 2014, \prl, 112, 251302

\bibitem[{{Hellwing} {et~al.}(2021){Hellwing}, {Cautun}, {van de Weygaert}, \&
  {Jones}}]{hellwing2021}
{Hellwing}, W.~A., {Cautun}, M., {van de Weygaert}, R., \& {Jones}, B.~T. 2021,
  \prd, 103, 063517

\bibitem[{{Hoyle} \& {Vogeley}(2002)}]{2002ApJ...566..641H}
{Hoyle}, F. \& {Vogeley}, M.~S. 2002, \apj, 566, 641

\bibitem[{{Hoyle} \& {Vogeley}(2004)}]{2004ApJ...607..751H}
{Hoyle}, F. \& {Vogeley}, M.~S. 2004, \apj, 607, 751

\bibitem[{{Huchra} {et~al.}(2012){Huchra}, {Macri}, {Masters}, {Jarrett},
  {Berlind}, {Calkins}, {Crook}, {Cutri}, {Erdo{\v{g}}du}, {Falco}, {George},
  {Hutcheson}, {Lahav}, {Mader}, {Mink}, {Martimbeau}, {Schneider},
  {Skrutskie}, {Tokarz}, \& {Westover}}]{huchra2012}
{Huchra}, J.~P., {Macri}, L.~M., {Masters}, K.~L., {et~al.} 2012, \apjs, 199,
  26

\bibitem[{{Icke}(1984)}]{icke1984}
{Icke}, V. 1984, \mnras, 206, 1P

\bibitem[{{Kirshner} {et~al.}(1981){Kirshner}, {Oemler}, {Schechter}, \&
  {Shectman}}]{kirshner1981}
{Kirshner}, R.~P., {Oemler}, A., J., {Schechter}, P.~L., \& {Shectman}, S.~A.
  1981, \apjl, 248, L57

\bibitem[{{Kreckel} {et~al.}(2012){Kreckel}, {Platen}, {Arag{\'o}n-Calvo}, {van
  Gorkom}, {van de Weygaert}, {van der Hulst}, \& {Beygu}}]{kreckel2012}
{Kreckel}, K., {Platen}, E., {Arag{\'o}n-Calvo}, M.~A., {et~al.} 2012, \aj,
  144, 16

\bibitem[{{Kreckel} {et~al.}(2011){Kreckel}, {Platen}, {Arag{\'o}n-Calvo}, {van
  Gorkom}, {van de Weygaert}, {van der Hulst}, {Kova{\v{c}}}, {Yip}, \&
  {Peebles}}]{kreckel2011}
{Kreckel}, K., {Platen}, E., {Arag{\'o}n-Calvo}, M.~A., {et~al.} 2011, \aj,
  141, 4

\bibitem[{{Lackner} {et~al.}(2012){Lackner}, {Cen}, {Ostriker}, \&
  {Joung}}]{lackner2012}
{Lackner}, C.~N., {Cen}, R., {Ostriker}, J.~P., \& {Joung}, M.~R. 2012, \mnras,
  425, 641

\bibitem[{{Lavaux} \& {Wandelt}(2010)}]{lavaux2010}
{Lavaux}, G. \& {Wandelt}, B.~D. 2010, \mnras, 403, 1392

\bibitem[{{Lavaux} \& {Wandelt}(2012)}]{lavaux2011}
{Lavaux}, G. \& {Wandelt}, B.~D. 2012, \apj, 754, 109

\bibitem[{{Pan} {et~al.}(2012){Pan}, {Vogeley}, {Hoyle}, {Choi}, \&
  {Park}}]{2012MNRAS.421..926P}
{Pan}, D.~C., {Vogeley}, M.~S., {Hoyle}, F., {Choi}, Y.-Y., \& {Park}, C. 2012,
  \mnras, 421, 926

\bibitem[{{Paranjape} {et~al.}(2018){Paranjape}, {Hahn}, \&
  {Sheth}}]{paranjape2018}
{Paranjape}, A., {Hahn}, O., \& {Sheth}, R.~K. 2018, \mnras, 476, 3631

\bibitem[{{Park} \& {Lee}(2007)}]{parklee2007}
{Park}, D. \& {Lee}, J. 2007, \prl, 98, 081301

\bibitem[{{Paz} {et~al.}(2013){Paz}, {Lares}, {Ceccarelli}, {Padilla}, \&
  {Lambas}}]{paz2013}
{Paz}, D., {Lares}, M., {Ceccarelli}, L., {Padilla}, N., \& {Lambas}, D.~G.
  2013, \mnras, 436, 3480

\bibitem[{{Peebles}(1980)}]{peebles1980}
{Peebles}, P.~J.~E. 1980, {The large-scale structure of the universe}

\bibitem[{{Peebles}(2001)}]{peebles2001}
{Peebles}, P.~J.~E. 2001, \apj, 557, 495

\bibitem[{{P\'erez} \& {al.}(2023)}]{Perez2023}
{P\'erez}, I. \& {al.} 2023 [\eprint{in preparation}]

\bibitem[{{Perico} {et~al.}(2019){Perico}, {Voivodic}, {Lima}, \&
  {Mota}}]{perico2019}
{Perico}, E. L.~D., {Voivodic}, R., {Lima}, M., \& {Mota}, D.~F. 2019, \aap,
  632, A52

\bibitem[{{Pisani} {et~al.}(2019){Pisani}, {Massara}, {Spergel}, {Alonso},
  {Baker}, {Cai}, {Cautun}, {Davies}, {Demchenko}, {Dor{\'e}}, {Goulding},
  {Habouzit}, {Hamaus}, {Hawken}, {Hirata}, {Ho}, {Jain}, {Kreisch}, {Marulli},
  {Padilla}, {Pollina}, {Sahl{\'e}n}, {Sheth}, {Somerville}, {Szapudi}, {van de
  Weygaert}, {Villaescusa-Navarro}, {Wandelt}, \& {Wang}}]{pisani2020}
{Pisani}, A., {Massara}, E., {Spergel}, D.~N., {et~al.} 2019, \baas, 51, 40

\bibitem[{{Pisani} {et~al.}(2015){Pisani}, {Sutter}, {Hamaus}, {Alizadeh},
  {Biswas}, {Wandelt}, \& {Hirata}}]{pisani2015}
{Pisani}, A., {Sutter}, P.~M., {Hamaus}, N., {et~al.} 2015, \prd, 92, 083531

\bibitem[{{Planck Collaboration} {et~al.}(2016){Planck Collaboration}, {Ade},
  {Aghanim}, {Arnaud}, {Ashdown}, {Aumont}, {Baccigalupi}, {Banday},
  {Barreiro}, {Bartlett}, {Bartolo}, {Battaner}, {Battye}, {Benabed},
  {Beno{\^\i}t}, {Benoit-L{\'e}vy}, {Bernard}, {Bersanelli}, {Bielewicz},
  {Bock}, {Bonaldi}, {Bonavera}, {Bond}, {Borrill}, {Bouchet}, {Boulanger},
  {Bucher}, {Burigana}, {Butler}, {Calabrese}, {Cardoso}, {Catalano},
  {Challinor}, {Chamballu}, {Chary}, {Chiang}, {Chluba}, {Christensen},
  {Church}, {Clements}, {Colombi}, {Colombo}, {Combet}, {Coulais}, {Crill},
  {Curto}, {Cuttaia}, {Danese}, {Davies}, {Davis}, {de Bernardis}, {de Rosa},
  {de Zotti}, {Delabrouille}, {D{\'e}sert}, {Di Valentino}, {Dickinson},
  {Diego}, {Dolag}, {Dole}, {Donzelli}, {Dor{\'e}}, {Douspis}, {Ducout},
  {Dunkley}, {Dupac}, {Efstathiou}, {Elsner}, {En{\ss}lin}, {Eriksen},
  {Farhang}, {Fergusson}, {Finelli}, {Forni}, {Frailis}, {Fraisse},
  {Franceschi}, {Frejsel}, {Galeotta}, {Galli}, {Ganga}, {Gauthier}, {Gerbino},
  {Ghosh}, {Giard}, {Giraud-H{\'e}raud}, {Giusarma}, {Gjerl{\o}w},
  {Gonz{\'a}lez-Nuevo}, {G{\'o}rski}, {Gratton}, {Gregorio}, {Gruppuso},
  {Gudmundsson}, {Hamann}, {Hansen}, {Hanson}, {Harrison}, {Helou},
  {Henrot-Versill{\'e}}, {Hern{\'a}ndez-Monteagudo}, {Herranz}, {Hildebrandt},
  {Hivon}, {Hobson}, {Holmes}, {Hornstrup}, {Hovest}, {Huang}, {Huffenberger},
  {Hurier}, {Jaffe}, {Jaffe}, {Jones}, {Juvela}, {Keih{\"a}nen}, {Keskitalo},
  {Kisner}, {Kneissl}, {Knoche}, {Knox}, {Kunz}, {Kurki-Suonio}, {Lagache},
  {L{\"a}hteenm{\"a}ki}, {Lamarre}, {Lasenby}, {Lattanzi}, {Lawrence}, {Leahy},
  {Leonardi}, {Lesgourgues}, {Levrier}, {Lewis}, {Liguori}, {Lilje},
  {Linden-V{\o}rnle}, {L{\'o}pez-Caniego}, {Lubin}, {Mac{\'\i}as-P{\'e}rez},
  {Maggio}, {Maino}, {Mandolesi}, {Mangilli}, {Marchini}, {Maris}, {Martin},
  {Martinelli}, {Mart{\'\i}nez-Gonz{\'a}lez}, {Masi}, {Matarrese}, {McGehee},
  {Meinhold}, {Melchiorri}, {Melin}, {Mendes}, {Mennella}, {Migliaccio},
  {Millea}, {Mitra}, {Miville-Desch{\^e}nes}, {Moneti}, {Montier}, {Morgante},
  {Mortlock}, {Moss}, {Munshi}, {Murphy}, {Naselsky}, {Nati}, {Natoli},
  {Netterfield}, {N{\o}rgaard-Nielsen}, {Noviello}, {Novikov}, {Novikov},
  {Oxborrow}, {Paci}, {Pagano}, {Pajot}, {Paladini}, {Paoletti}, {Partridge},
  {Pasian}, {Patanchon}, {Pearson}, {Perdereau}, {Perotto}, {Perrotta},
  {Pettorino}, {Piacentini}, {Piat}, {Pierpaoli}, {Pietrobon}, {Plaszczynski},
  {Pointecouteau}, {Polenta}, {Popa}, {Pratt}, {Pr{\'e}zeau}, {Prunet},
  {Puget}, {Rachen}, {Reach}, {Rebolo}, {Reinecke}, {Remazeilles}, {Renault},
  {Renzi}, {Ristorcelli}, {Rocha}, {Rosset}, {Rossetti}, {Roudier},
  {Rouill{\'e} d'Orfeuil}, {Rowan-Robinson}, {Rubi{\~n}o-Mart{\'\i}n},
  {Rusholme}, {Said}, {Salvatelli}, {Salvati}, {Sandri}, {Santos},
  {Savelainen}, {Savini}, {Scott}, {Seiffert}, {Serra}, {Shellard}, {Spencer},
  {Spinelli}, {Stolyarov}, {Stompor}, {Sudiwala}, {Sunyaev}, {Sutton},
  {Suur-Uski}, {Sygnet}, {Tauber}, {Terenzi}, {Toffolatti}, {Tomasi},
  {Tristram}, {Trombetti}, {Tucci}, {Tuovinen}, {T{\"u}rler}, {Umana},
  {Valenziano}, {Valiviita}, {Van Tent}, {Vielva}, {Villa}, {Wade}, {Wandelt},
  {Wehus}, {White}, {White}, {Wilkinson}, {Yvon}, {Zacchei}, \&
  {Zonca}}]{Planck2015}
{Planck Collaboration}, {Ade}, P.~A.~R., {Aghanim}, N., {et~al.} 2016, \aap,
  594, A13

\bibitem[{{Platen} {et~al.}(2007){Platen}, {van de Weygaert}, \&
  {Jones}}]{platen2007}
{Platen}, E., {van de Weygaert}, R., \& {Jones}, B. J.~T. 2007, \mnras, 380,
  551

\bibitem[{{Pomar{\`e}de} {et~al.}(2017){Pomar{\`e}de}, {Hoffman}, {Courtois},
  \& {Tully}}]{2017ApJ...845...55P}
{Pomar{\`e}de}, D., {Hoffman}, Y., {Courtois}, H.~M., \& {Tully}, R.~B. 2017,
  \apj, 845, 55

\bibitem[{{Ricciardelli} {et~al.}(2013){Ricciardelli}, {Quilis}, \&
  {Planelles}}]{ricciardelli2013}
{Ricciardelli}, E., {Quilis}, V., \& {Planelles}, S. 2013, \mnras, 434, 1192

\bibitem[{{Rojas} {et~al.}(2004){Rojas}, {Vogeley}, {Hoyle}, \&
  {Brinkmann}}]{rojas2004}
{Rojas}, R.~R., {Vogeley}, M.~S., {Hoyle}, F., \& {Brinkmann}, J. 2004, \apj,
  617, 50

\bibitem[{{Rojas} {et~al.}(2005){Rojas}, {Vogeley}, {Hoyle}, \&
  {Brinkmann}}]{rojas2005}
{Rojas}, R.~R., {Vogeley}, M.~S., {Hoyle}, F., \& {Brinkmann}, J. 2005, \apj,
  624, 571

\bibitem[{{Roth} {et~al.}(2005){Roth}, {Kelz}, {Fechner}, {Hahn}, {Bauer},
  {Becker}, {B{\"o}hm}, {Christensen}, {Dionies}, {Paschke}, {Popow}, {Wolter},
  {Schmoll}, {Laux}, \& {Altmann}}]{2005PASP..117..620R}
{Roth}, M.~M., {Kelz}, A., {Fechner}, T., {et~al.} 2005, \pasp, 117, 620

\bibitem[{{Sheth} \& {van de Weygaert}(2004)}]{shethwey2004}
{Sheth}, R.~K. \& {van de Weygaert}, R. 2004, \mnras, 350, 517

\bibitem[{{Tinker} {et~al.}(2006){Tinker}, {Weinberg}, \&
  {Warren}}]{tinker2006}
{Tinker}, J.~L., {Weinberg}, D.~H., \& {Warren}, M.~S. 2006, \apj, 647, 737

\bibitem[{{Tully} {et~al.}(2014){Tully}, {Courtois}, {Hoffman}, \&
  {Pomar{\`e}de}}]{tully2014}
{Tully}, R.~B., {Courtois}, H., {Hoffman}, Y., \& {Pomar{\`e}de}, D. 2014,
  \nat, 513, 71

\bibitem[{{Tully} {et~al.}(2016){Tully}, {Courtois}, \&
  {Sorce}}]{2016AJ....152...50T}
{Tully}, R.~B., {Courtois}, H.~M., \& {Sorce}, J.~G. 2016, \aj, 152, 50

\bibitem[{{Tully} {et~al.}(2023){Tully}, {Kourkchi}, {Courtois}, {Anand},
  {Blakeslee}, {Brout}, {Jaeger}, {Dupuy}, {Guinet}, {Howlett}, {Jensen},
  {Pomar{\`e}de}, {Rizzi}, {Rubin}, {Said}, {Scolnic}, \&
  {Stahl}}]{2023ApJ...944...94T}
{Tully}, R.~B., {Kourkchi}, E., {Courtois}, H.~M., {et~al.} 2023, \apj, 944, 94

\bibitem[{{Vall{\'e}s-P{\'e}rez} {et~al.}(2021){Vall{\'e}s-P{\'e}rez},
  {Quilis}, \& {Planelles}}]{2021ApJ...920L...2V}
{Vall{\'e}s-P{\'e}rez}, D., {Quilis}, V., \& {Planelles}, S. 2021, \apjl, 920,
  L2

\bibitem[{{van de Weygaert}(2016)}]{weygaert2016}
{van de Weygaert}, R. 2016, in The Zeldovich Universe: Genesis and Growth of
  the Cosmic Web, ed. R.~{van de Weygaert}, S.~{Shandarin}, E.~{Saar}, \&
  J.~{Einasto}, Vol. 308, 493--523

\bibitem[{{van de Weygaert} \& {Babul}(1994)}]{weybab1994}
{van de Weygaert}, R. \& {Babul}, A. 1994, \apjl, 425, L59

\bibitem[{{van de Weygaert} \& {Platen}(2011)}]{weygaert2011}
{van de Weygaert}, R. \& {Platen}, E. 2011, in International Journal of Modern
  Physics Conference Series, Vol.~1, International Journal of Modern Physics
  Conference Series, 41--66

\bibitem[{{van de Weygaert} \& {van Kampen}(1993)}]{weykamp1993}
{van de Weygaert}, R. \& {van Kampen}, E. 1993, \mnras, 263, 481

\bibitem[{{Verza} {et~al.}(2022){Verza}, {Carbone}, \& {Renzi}}]{verza2022}
{Verza}, G., {Carbone}, C., \& {Renzi}, A. 2022, \apjl, 940, L16

\bibitem[{{Yan} {et~al.}(2013){Yan}, {Fan}, \& {White}}]{yan2013}
{Yan}, H., {Fan}, Z., \& {White}, S. D.~M. 2013, \mnras, 430, 3432

\end{thebibliography}
\bibliographystyle{aa}

%\clearpage
\appendix

\section{Void density and velocity profiles}
\label{app:appendixA}

One may analytically compute the expected density and velocity profiles of isolated spherical voids, into
the far nonlinear regime, up to the moment that voids experience shell crossing at their boundaries \citep[for a review see][]{weygaert2016}.
The explicit expression for the density and velocity profiles for an isolated spherical void may be found in \cite{shethwey2004}.

One major result is that voids have a characteristic density and evolution time, that of shell crossing.
For a spherical void, this happens when it reaches a nonlinear density contrast $\delta \sim -0.8$ (i.e., $20\%$ of
the global cosmic density), by which time the void has expanded by a factor $\sim 1.7$. This corresponds
to a linear density contrast of $\delta_{lin} =- 2.81$ (this is to be compared to the 1.69 for collapse of
spherical overdensities).

\subsection{Void density profiles}
We may also use the expansion of isolated spherical voids to understand the overall density and velocity profiles
of voids  \cite[see also][]{weygaert2016} because they are underdense, they expand with respect to the background, and the
interior shells expand faster than the outer ones. Due to the differential expansion of the interior mass shells, we
see an accumulation of mass near the exterior and boundary of the void, meanwhile evening out the density distribution
in the interior. This leads to a typical {bucket-shaped} density profile (opposite of top-hat),
with a linear "Hubble-like" void flow in the interior (the canonical void is a "Hubble bubble").
For a wide range of initial radial profiles, voids will attain a bucket-shaped profile.

Recently, a range of studies have been published on the issue of void-density profiles, and the question 
of whether or not they display a universal behavior \citep[see e.g.,][]{hamaus2014,cautun2016}. For  example, \citet[][]{ricciardelli2013} and \citet{hamaus2014} concluded that spherically averaged density profiles 
of voids indeed imply a universal density profile that can be characterized by two parameters.

Interestingly, these density profiles have a less prominent bucket-shaped interior profile than 
those seen for the spherical voids. This may be understood from the 
fact that voids in general are not spherical, meaning that spherical averaging 
will lead to the mixing of different layers in the  interior of a void. The recent study by \cite{cautun2016} 
confirms this: when taking into account the shapes of voids, a remarkably strong bucket 
void density profile appears. 

\subsection{Void velocity profiles}
In the situation of a mature, evolved void, the velocity field of a void resembles that 
of a Hubble flow, in which the outflow velocity increases linearly with distance to the 
void center. In other words, voids are {super-Hubble bubbles} \citep{icke1984}.
The linear velocity increase is a reflection of the corresponding density distribution: the near constant 
velocity divergence within the void conforms to the near uniform {bucket-shaped} interior density distribution
that voids attain at more advanced stages.

It is straightforward to appreciate this from the {continuity equation}. For a uniform 
density field, this equation tells us that the velocity divergence in the void will be uniform, 
corresponding to a Hubble-like outflow. Because voids are emptier than the rest of 
the Universe, they will expand faster, with a net 
velocity divergence equal to
\begin{eqnarray}
   \theta&\,=\,&{\displaystyle \nabla\cdot{\bf v} \over \displaystyle H}\,=\,3 (\alpha-1)\,,\qquad\alpha=H_{\rm void}/H\,,
\end{eqnarray}
\noindent where $\alpha$ is defined to be the ratio of the super-Hubble expansion rate of the 
void and the Hubble expansion of the Universe.  \cite{weykamp1993} confirmed that the velocity outflow field in
viable cosmological scenarios does indeed resemble that of a super-Hubble expanding bubble. 
These authors established that the super-Hubble expansion rate is directly proportional to the nonlinear 
void density $\Delta(t)$, 
\begin{equation}
H_{\rm void}/H\,=\,-{\displaystyle 1 \over \displaystyle 3} f(\Omega)\,\Delta(t)\,.
\end{equation}
This relation, known within the context of a linearly evolving spherical density perturbation, 
in the case of fully evolved voids appears to be valid on the basis of the {\it nonlinear} 
void density deficit. Several studies \citep[e.g.,][]{hamaus2014} have 
confirmed this finding for voids in a range of high-resolution cosmological simulations.
The immediate implication is that voids should be considered 
as distinctly {\it nonlinear} objects. 

\subsection{Nonspherical voids}

One may wonder how far the nonsphericity of voids works out for their density and velocity profiles.
\cite{cautun2016} showed that this may severely influence the density and velocity profiles
extracted for these voids and that spherical averaging may not always lead to a correct result.
In the interpretation of the profiles presented in the main text, these effects will certainly play a role.

\bigskip
While the results above emanate from a rather unrealistic symmetric
configuration ---spherical, isolated---  many studies have shown these to be rather representative
for the major fully expanding voids in the galaxy distribution.
There is also a good reason why this is so: following the shell-crossing phase, the expansion of voids
slows down \citep{bertschinger1985}. It was this realization that led \cite{dubinski1992} to point out
that the large voids in redshift surveys are to be mostly identified with the voids that at the current epoch are undergoing shell crossing.

\end{document}